\documentclass[%
 amsmath,amssymb,
 reprint,%
]{revtex4-1}

\usepackage{graphicx}
\usepackage{dcolumn}
\usepackage{bm}

\usepackage[utf8]{inputenc}
\usepackage[T1]{fontenc}
\usepackage{mathptmx}
\usepackage{etoolbox}
\usepackage{verbatim}
\usepackage{hyperref}
\usepackage{bbold}
\usepackage{xcolor}
\usepackage{mathrsfs}

\usepackage{color} 
\usepackage{hyperref}
\usepackage{amsmath,amssymb,mathrsfs}
\usepackage{psfrag}
\usepackage{graphicx}
\usepackage{graphics}
\usepackage{subfigure}
\usepackage{epsfig}
\usepackage{bm}
\usepackage{verbatim,color}
\usepackage{braket}
\usepackage[USenglish]{babel}
\usepackage{lipsum}
\usepackage[normalem]{ulem}

\makeatletter
\def\@email#1#2{%
 \endgroup
 \patchcmd{\titleblock@produce}
  {\frontmatter@RRAPformat}
  {\frontmatter@RRAPformat{\produce@RRAP{*#1\href{mailto:#2}{#2}}}\frontmatter@RRAPformat}
  {}{}
}%
\makeatother
\begin{document}

\preprint{AIP/123-QED}

\title[]{Environment-limited transfer of angular momentum in Bose liquids}

\author{Alberto Cappellaro}
 \email{alberto.cappellaro@ist.ac.at}
 \affiliation{Institute of Science and Technology Austria (ISTA), \\
 Am Campus 1, 3400 Klosterneuburg, Austria}

\author{Giacomo Bighin}
 \affiliation{Institute of Science and Technology Austria (ISTA), \\
 Am Campus 1, 3400 Klosterneuburg, Austria}

\author{Igor Cherepanov}
 \affiliation{Institute of Science and Technology Austria (ISTA), \\
 Am Campus 1, 3400 Klosterneuburg, Austria}
 
\author{Mikhail Lemeshko}%
 \affiliation{Institute of Science and Technology Austria (ISTA), \\
 Am Campus 1, 3400 Klosterneuburg, Austria}

\date{\today}

\begin{abstract}
Impurity motion in a many-body environment has been a central issue in the field of 
low-temperature physics for decades. 
In bosonic quantum fluids, the onset of a drag force
experienced by point-like objects is due to collective environment excitations, driven by 
the exchange of linear momentum between the impurity and the many-body bath. 
In this work we consider a rotating impurity, with the aim of exploring how angular momentum
is exchanged with the surrounding bosonic environment. In order to elucidate this issues, we 
employ a quasiparticle approach based on the angulon theory, which allows us to effectively
deal with the non-trivial algebra of quantized angular momentum in presence of
a many-body environment. We uncover how impurity dressing by 
environmental excitations can establish an exchange channel,
whose effectiveness crucially depends on the initial state of the impurity. Remarkably, we find that
there is a critical value of initial angular momentum, above which this channel effectively freezes.
\end{abstract}

\maketitle

\section{Introduction and motivations}
Superfluidity has been a central topic in the field of low-temperature physics since the 
first half of the last century, after Kapitza, Allen and Misener provided a firm experimental
probe of viscosity suppression in liquid helium below the $\lambda$-point in 1938
 \cite{Kapitza-1938,Allen-1938}. 
Remarkably, much of superfluidity phenomenology (especially the transport-related issues) are
effectively captured by Landau's phenomenological approach \cite{landau-1941,annett-2004,pitaevskii-2016}. 
By introducing the 
concept of quasiparticle excitations, the resulting picture reads 
a threshold velocity $v_L$ below which 
a (linearly) moving object does not experience any 
dissipation. The only required input is the excitation spectrum $E(k)$, such that
$v_L = \min \big[ E(k)/ (\hbar k)\big]$.
Landau's phenomenological approach can be formalized by invoking Galilean invariance,
unveiling the concept of superfluid (or ground state) rigidity, where a small momentum boost
is enough to set the superfluid in motion with constant velocity.
While Landau provided no microscopic picture, this scenario can be substantiated by the
emergence of a (complex) macroscopic wavefunction \cite{Baym-1968}, 
whose phase $\theta$ is related to the so-called superfluid
velocity, defined as $\mathbf{v}_s = \hbar\nabla \theta /m$. 

Beyond Landau's paradigm, additional insight is provided by microscopic calculations
of the drag force acting on an impurity moving through the superfluid bulk
\cite{roberts-2005,roberts-2006,brand-2011}.
By performing a 
Galilean transformation to the frame of the impurity, one gets
$ \mathbf{F} = - \int d^3\mathbf{r} \langle 
\hat{\psi}^{\dagger}(\mathbf{r}) [\eta \nabla \delta(\mathbf{r})] \hat{\psi}(\mathbf{r})
\rangle $,
where $\hat{\psi}$ is the field operator describing the ordered moving background,
and $\eta$ being the impurity-environment coupling strength. The structureless
character of the impurity is accounted for by the delta potential. Mean-field calculations,
based on the assumption of a perfect condensate, agrees with Landau's picture: 
a drag force is experienced only if 
the impurity moves faster than $v_L$ \cite{astrakharcik-2004}. 
On the other hand, when fluctuations are included, it can be shown that the 
Landau velocity is no more a
sharp boundary between the frictionless and dissipative flows. Instead, for $v < v_L$ 
the drag force grows linearly with the Mach number 
$v/c_s$, $c_s$ being the sound velocity \cite{suzuki-2014}. 
%
%
In a nutshell, the onset of a drag force is driven by the exchange of linear
momentum between the impurity and the superfluid environment \cite{brand-2011}. 

The picture definitely becomes much more involved when we move from point-like objects,
to impurities with a richer structure, where the geometric arrangement of their
constituents endow them with the ability to perform rotations in real space. 
Rotational motion of extended objects was employed in the past to probe the onset of 
superfluidity, for instance by measuring how the inertia moment of a torsion pendulum changes when 
crossing $^4$He $\lambda$-point \cite{andronikashvili-1966,reppy-1978}. However, in that case,
rotational degrees of freedom were treated classically, 
thus avoiding the complications arising from the
non-trivial algebra of quantized angular momentum. 
Such treatment cannot be avoided in the case of molecules: quantum 
control on rotational dynamics
has become increasingly relevant in the field of cold chemistry\cite{koch-2019,krems-book}, since
reactivity also depends on the relative orientation of the molecules involved. 
A similar argument can be made for nanoparticles in the relatively new field of levitodynamics
\cite{Stickler-2021,levitodynamics-review}.
Now, on the molecular side, a paradigmatic quantum environment is provided by superfluid $^4$He;
molecules have been routinely trapped in helium nanodroplets for more than two decades, 
mostly for spectroscopic application\cite{toennies-1998,toennies-2013}. 
While spectroscopic lines are not broadened by the environment, 
it has been shown early how superfluid helium can affect 
the moments of inertia, a feature particularly pronounced for heavy species\cite{toennies-2004} 
like  CS$_2$. On the theory side, molecular rotations in superfluids have been investigated
through different and refined numerical methods, such as path-integral and diffusion quantum 
MonteCarlo \cite{ceperley-1996,zillich-2005,miura-2007-1,miura-2007-2,villareal-2016}, together
with density functional and hydrodynamical approaches \cite{dalfovo-1999,ancilotto-2017}. 

In this paper, we focus on the issue of angular momentum transfer between a rotating impurity and the
surrounding bosonic environment. In order to tackle down this problem, we make use of an alternative
approach based on the so-called \textit{angulon} theory
\cite{lemeshko-2015, lemeshko-2016, lemeshko-2017,lemeshko-2017-prl}. 
This framework is built upon the formalism of \textit{impurity physics}, originally developed
in solid-state physics where, for instance, translational motion of an electron across an ionic 
lattice is often referred to as the \textit{polaron problem}, a concept which has been proved
extremely useful even in the context of atomic BECs \cite{novikov-2010,shchadilova-2016}. 
On the other hand, the angulon 
describes how a rotating impurity is affected by the excitations of a many-body environment. 
In order to fulfill this task, the non-Abelian $\text{SO}(3)$ algebra of quantized angular momentum has to be
fully encoded in the theory. In principle, this is a highly non-trivial task, since a 
many-body environment may entail the necessity to include an infinite number of 
angular momentum vectors, whose coupling introduces Wigner 3$jn$ coefficients of
arbitrarily high orders, resulting in an algebraically intractable problem
\cite{varshalovitch-book}. This technical problem is properly addressed in the angulon framework, 
showing how the dressing mechanism is established, leading up to an open channel to exchange 
angular momentum with the environment.

For the sake of clarity, in the following we are considering a rigid linear quantum 
rotor, a well-tested approximation for linear molecules in the $^1\Sigma$ electronic state
\cite{krems-book}. 
%
%
By employing a variational approach, we are able to show that angular momentum transfer dynamics is 
oscillatory, with the relaxation time being drastically dependent on the initial state of the impurity.
The onset mechanism of this dynamical features is related to environment 
excitations effectively 
dressing the rotor. This mechanism is in place for low values of initial angular momentum,
while at higher values  medium fluctuations heavily perturb the impurity, \textit{diluting} its 
its rotating character in such a way that the exchange channel for angular momentum is effectively closed. 

%
%
%
%
%
%

\section{The angulon framework}
\label{sec:angulon}
Here we focus on the
Hamiltonian $\hat{H}_A$, describing the whole system made of a rotating object 
(here, a linear rigid rotor) and the 
(superfluid) environment in which it is immersed. As mentioned in the introduction, 
a convenient description for the rotational dynamics is given in terms of the 
angulon quasiparticle\cite{lemeshko-2015,lemeshko-2016}. We also assume the linear motion to be 
frozen, with no loss of generality, as it is the case for many experimentally relevant cases
\cite{toennies-2004}.
Indeed, in the
laboratory frame one gets ($\hbar = 1$)
\begin{equation}
\begin{aligned} 
\hat{H}_A & = B \hat{\mathbf{J}}^2  + \sum_{k\lambda\mu} \omega_B(k) 
\hat{b}^{\dagger}_{k\lambda\mu} \hat{b}_{k\lambda\mu} \\ 
& \quad + \sum_{k\lambda\mu} U_{\lambda}(k)\bigg[ 
\hat{Y}_{\lambda\mu}^*(\hat{\theta},\hat{\phi})\,\hat{b}^{\dagger}_{k\lambda\mu}
+\hat{Y}_{\lambda\mu}(\hat{\theta},\hat{\phi})\,\hat{b}_{k\lambda\mu}
\bigg]\;.
\end{aligned}
\label{angulon hamiltonian lab frame}
\end{equation}
where $k = |\mathbf{k}|$ and $\sum_k \equiv \int dk$, with $\lambda$ and $\mu$ being
respectively the angular momentum and its projection on the $\hat{z}$-axis in the laboratory
frame. 
The bosonic bath operators are cast 
in the spherical representation \cite{greiner-book}, with the usual commutation relation
$\big[\hat{b}_{k\lambda\mu}, \hat{b}_{k'\lambda'\mu'}^{\dagger}\big] 
= \delta(k - k')\delta_{\lambda \lambda' \mu \mu'}$. 
In the equation above, $\omega_B(k)$ is the dispersion relation of the bath excitations,
while the impurity-bath coupling is provided by
\begin{equation}
U_{\lambda}(k) = u_{\lambda}\sqrt{\frac{8 n k^2 \epsilon(k)}{\omega_B(k)(2\lambda +1)}}
\int dr\, r^2  j_{\lambda}(kr) \, V_{\lambda}(r)\;,  
\label{shape of impurity-bath coupling}
\end{equation}
with $\epsilon_k = k^2/2m$, $n$ the bath density and $V_{\lambda}(r)$ a properly defined 
potential shape in the corresponding $\lambda$-channel. 
Here, we assume a Gaussian form factor, i.e. 
$V_{\lambda}(r) = (2\pi)^{3/2} \, e^{-r^2/2r_{\lambda}^2}$.

\begin{figure}[ht!]
\includegraphics[width=0.80\columnwidth]{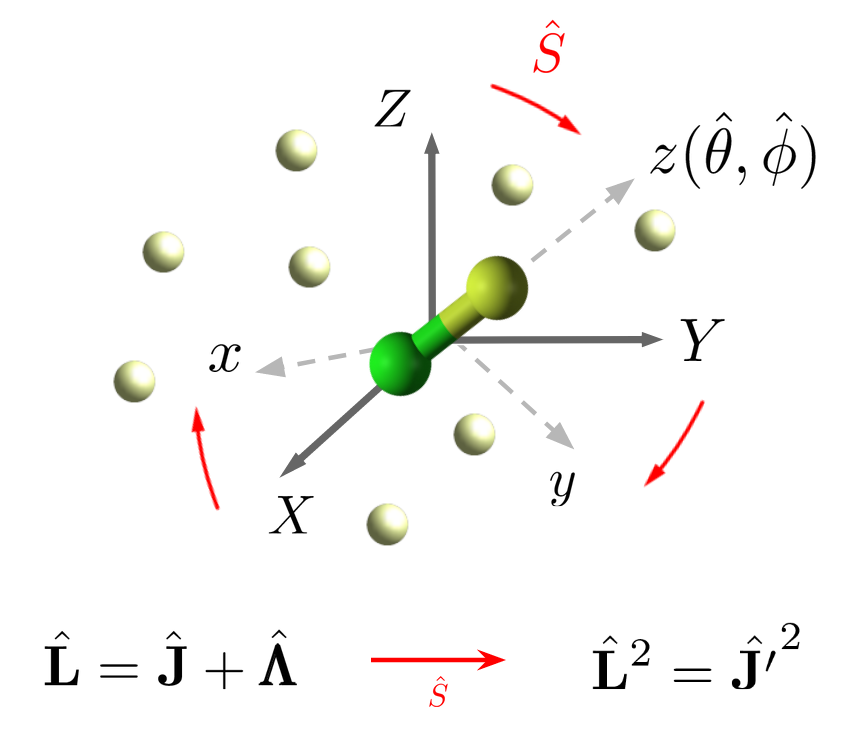}
\caption{Representation of the theoretical framework described in Sec. \ref{sec:angulon}. A diatomic molecule,
modelled as a linear rigid rotor, is immersed in a bosonic bath; the molecule-boson coupling is introduced, in the laboratory frame, by the hamiltonian $\hat{H}$ in Eq. \eqref{angulon hamiltonian lab frame}. In
this frame (defined by the triad $\lbrace X, Y, Z\rbrace$), the total angular momentum $\hat{\mathbf{L}}$
is composed by two different contributions, $\hat{\mathbf{J}}$ accounting for the rotating impurity and
$\hat{\pmb{\Lambda}}$ for the bath (cf. Eq. \eqref{angular momentum of the bath}). On the other hand, when
the unitary transformation defined by $\hat{S}$ (see Eq. \eqref{unitary transformation matrix angles}) is
applied, environmental degrees of freedom are transferred to the rotating frame (labelled by
$\lbrace x,y,z\rbrace$). Remarkably, in the rotating molecular frame, it can be proved that \cite{lemeshko-2016}
 $\hat{\mathbf{L}} = \hat{\mathbf{J}'}^2$, i.e. the transformed molecular angular momentum exactly
 coincides with the total one.}
\label{fig:1}
\end{figure}
%

%
Now, 
strictly speaking, Eqs. \eqref{angulon hamiltonian lab frame} 
and \eqref{shape of impurity-bath coupling}
are rigorously derived in the
the context of Bose-Einstein condensates (BECs) made of weakly interacting Bose gases
\cite{lemeshko-2015}. There, 
$\omega_B(k)$ is the standard Bogoliubov spectrum accounting for phononic excitations. Nevertheless, 
Hamiltonians with similar structure can be effectively engineered to 
successfully investigate molecular impurities in helium nanodroplets 
\cite{lemeshko-2017-prl,shepperson-2017,cherepanov-2021}, 
Rydberg atoms in BECs \cite{schmidt-2016} and even electrons in solids 
\cite{katsnelson-2019,fahnle-2016}. 
In these more complex scenarios, there can be a wider range of possible environmental excitations 
(e.g. maxons and rotons in liquid helium), and fitting on experimental data or MonteCarlo simulations
may be required \cite{ceperley-1995,barenghi-1998}. Concerning the impurity-bath coupling,
and restricting to the case of molecules embedded in helium nanodroplets, a realistic 
atom-molecule interaction includes multiple terms emerging from an expansion in 
spherical harmonics, which
can be computed through extensive numerical quantum chemistry approaches \cite{szalewicz-2008}.
However, the structure of Eq. \eqref{angulon hamiltonian lab frame} still holds, since each of 
the above-mentioned terms generates a coupling between bare molecular states, mediated by 
environment excitations. While a simple analytical form for a $^4$He environment as the one in
Eq. \eqref{shape of impurity-bath coupling} does not reproduce quantitatively any realistic 
interaction, the relevant parameters can be chosen to match the magnitude and the range 
of these two-body potentials.
%

In this scenario the actual good quantum number is the total angular momentum of the
impurity-bath ensemble. Actually,   
this becomes much more transparent by moving to the molecular (i.e. fixed-body) 
frame\cite{lemeshko-2016,lemeshko-2017}.
This is implemented through a unitary transformation defined by
\begin{equation}
\hat{S}= \exp\lbrace -i \hat{\phi} \otimes \hat{\Lambda}_z \rbrace
\exp\lbrace -i \hat{\theta} \otimes \hat{\Lambda}_y \rbrace
\exp\lbrace -i \hat{\gamma} \otimes \hat{\Lambda}_z \rbrace
\label{unitary transformation matrix angles}
\end{equation}
where $\lbrace \hat{\phi}, \hat{\theta},\hat{\gamma}\rbrace$ are Euler angles specifying the orientation of 
one reference frame with respect to the other one, while 
\begin{equation}
\hat{\pmb{\Lambda}} = \sum_{k\lambda\mu\nu} \hat{b}_{k\lambda\mu}^{\dagger}
\pmb{\sigma}^{\lambda}_{\mu\nu} \hat{b}_{k\lambda\nu}
\label{angular momentum of the bath}
\end{equation}
is the bath angular momentum, $\pmb{\sigma}^{\lambda}_{\mu\nu}$ being the matrix vector enforcing
the proper algebra for the representation weighted by $\lambda$ (see App. \ref{app:1}
for an explicit definition of these matrices). We refer to Fig. \ref{fig:1}
for a pictorial representation of this theoretical procedure.
By applying Eq. \eqref{unitary transformation matrix angles} to $\hat{H}$
in Eq. \eqref{angulon hamiltonian lab frame}, we end up with
$\hat{\mathcal{H}}_{\text{mol}} \equiv \hat{S}^{-1} \hat{H}_A \hat{S}$
and
\begin{equation}
\begin{aligned}
\hat{\mathcal{H}}_{\text{mol}} & = B \big(\hat{\mathbf{J}}'-\hat{\pmb{\Lambda}}\big)^2 + \sum_{k\lambda\mu} \omega_B(k) 
\hat{b}^{\dagger}_{k\lambda\mu} \hat{b}_{k\lambda\mu}  \\
& \qquad\qquad  + \sum_{k\lambda} \mathcal{V}_{\lambda}(k)\big(\hat{b}^{\dagger}_{k\lambda 0} + \hat{b}_{k\lambda 0}\big).
\end{aligned}
\label{hamiltonian molecular frame}
\end{equation}
In the equation above $\hat{\mathbf{J}}'$ is the angular momentum operator acting on the molecular
frame, displaying an anomalous commutation relation 
$[\hat{J}_i',\hat{J}_j'] = -i\epsilon_{ijk}\hat{J}'_k$ \cite{bernath-book}.
Moreover, let us also notice that 
\begin{equation}
\mathcal{V}_{\lambda}(k) = U_{\lambda}(k)\sqrt{(2\lambda+1)/{4\pi}}\,
\label{slightly rescaled impurity-bath coupling}
\end{equation}
with $U_{\lambda}(k)$ being defined above in Eq. \eqref{shape of impurity-bath coupling}.

Thanks to the transformation described by $\hat{S}$ in 
Eq. \eqref{unitary transformation matrix angles}, 
environmental degrees of freedom are transferred 
to the rotating frame in such a way that the 
Euler angles operators are no more coupled to the excitation ladder operators.
Otherwise, Euler's angles (accounting for molecular orientation) should have been
coupled to every environmental excitation, rapidly resulting in an algebraically
intractable procedure.

\section{Discerning angular-momentum dynamics in the body-fixed frame} 
Concerning the emergent angular momentum dynamics in the bosonic environment, due to 
presence of the rotating impurity, we start by noticing that, due to the coupling, 
the actual conserved quantity is $\hat{\mathbf{L}} = \hat{\mathbf{J}}
+\hat{\pmb{\Lambda}}$, i.e. the total angular momentum. On the other hand, 
it can be shown \cite{lemeshko-2016} that $\hat{\mathcal{H}}_{\text{mol}}$ in 
the body-fixed frame is actually cast in terms of the total angular 
momentum, i.e. $\hat{\mathbf{L}}^2=\hat{\mathbf{J}}'^2$ 
\footnote{Equivalently, if $|L,M\rangle$ are eigenstates of $\hat{\mathbf{L}}^2$ 
and $\hat{L}_z$, the transformed states $\hat{S^{-1}}$ are also eigenstates
of $\hat{\mathcal{H}}_{\text{mol}}$. Check Ref. \cite{lemeshko-2016} for
a rigorous derivation of this property.}.
%
%
%
%
\begin{figure*}[t]
	\centering
	\begin{subfigure}
		\centering
        \includegraphics[width=\textwidth]{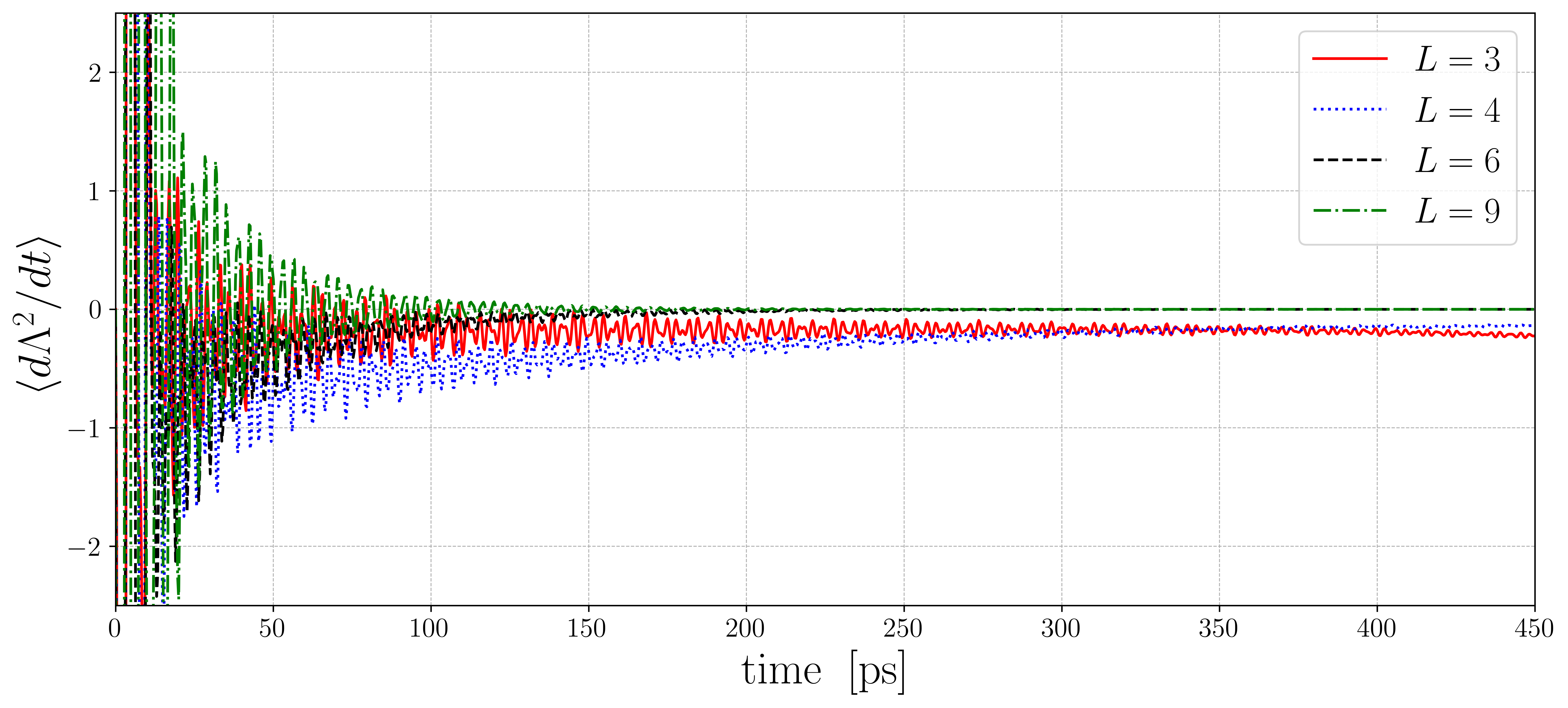}
		\label{fig:1a}
	\end{subfigure}
	\begin{subfigure}
		\centering
		\includegraphics[width=\textwidth]{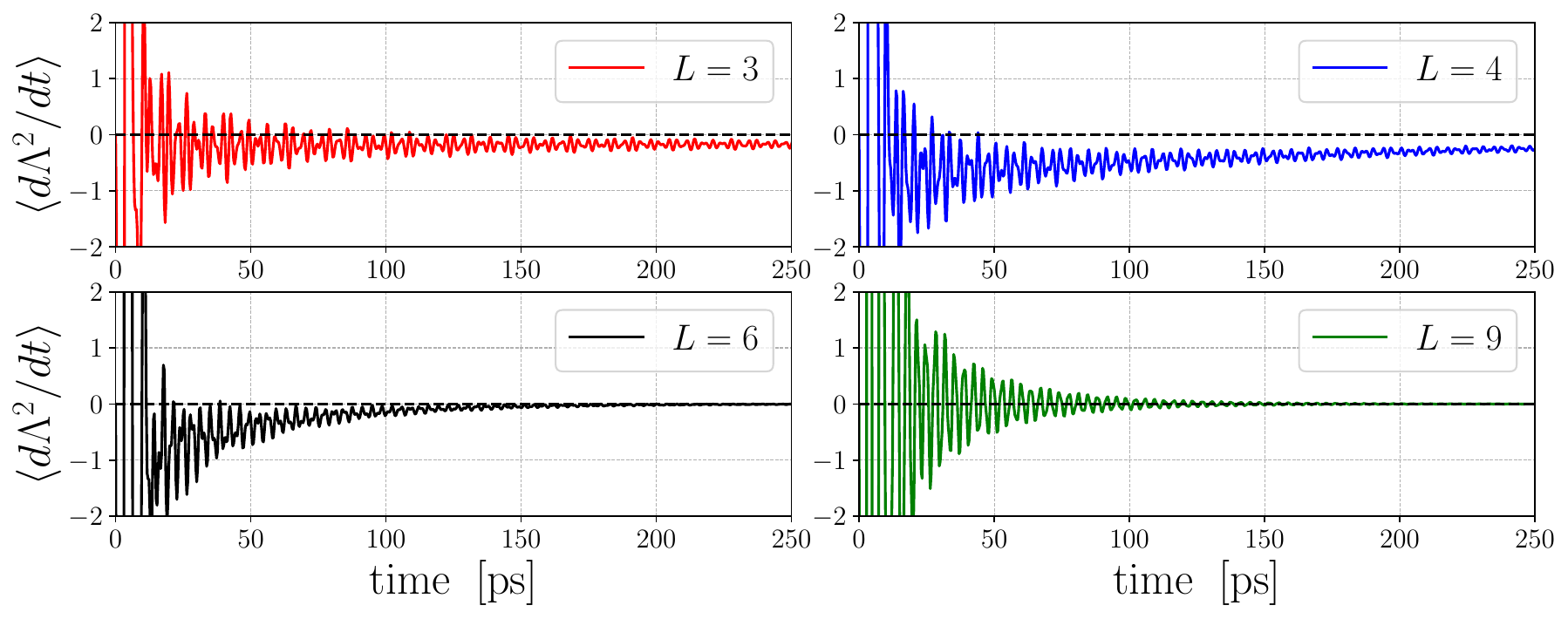}
        \label{fig:1b}
	\end{subfigure}
	\caption{\textit{Top panel}. 
		Time evolution of $\langle d\hat{\pmb{\Lambda}}^2/dt\rangle$
		at different values of the angular momentum $L$ in the initially prepared state 
		(i.e. $|\Psi(t=0)\rangle = | L, M = 0, n = 0\rangle$, with values of $L$ as in the panel
		and no phonons excited).  
		The numerical results displayed above have been obtained by solving the set of differential 
		equations derived within the time-dependent variational approach, where $\lambda  = 0, 2$ are the
		dominant terms for the molecule-bath interaction (see Eqs. \eqref{equation of motion for lambda zero two}
		in App. \ref{app:3}). 
        With the convention $\hbar = 1$, we adopt dimensionless units (denoted by a tilde), 
        such that energies are measured in units of $B$ 
        (the impurity rotational constants), distances are 
        multiples of $(mB)^{-1/2}$ and time is rescaled by $1/B$.
        We assume $B = 2\pi\hbar \times 1\; \mathrm{GHz} \simeq 0.03 \; \mathrm{cm}^{-1}$, while the
        other relevant parameters are the following: bath density is chosen $\tilde{n} = 74$, while 
        $\tilde{u}_0 = 218$, $\tilde{u}_2 = 320$ and 
        $\tilde{r}_0 = \tilde{r}_2 = 1.5$, respectively
		the magnitude of the molecule-bosons interaction and its characteristic range. 
        These parameters are chosen to reproduce the typical magnitude and range for a 
        homonuclear diatomic molecule such as, for instance, $I_2$.
        Results are sampled at a time step $\Delta \tilde{t} = 0.01$.  
		\textit{Bottom panels}. The same quantity considered 
        (at $L = 3, \;4,\; 6,\; 9$) for shorter times. 
        }
	\label{fig:2}
	\end{figure*}
As a consequence, in the body-fixed frame where the angular momentum
algebra is significantly simplified, we can explore the angular momentum transfer between 
the impurity and the environment by considering the time-evolution of 
$\hat{\pmb{\Lambda}}$ as in Eq. \eqref{angular momentum of the bath}. According to 
Ehrenfest theorem, we have
\begin{equation}
\bigg\langle \frac{d\hat{\pmb{\Lambda}}^2}{dt} \bigg\rangle
= -i \big\langle \big[\hat{\pmb{\Lambda}}^2,\hat{\mathcal{H}}_{\text{mol}}\big]\big \rangle
\label{lambda square ehrenfest}
\end{equation}
where we assumed that no external driving is present at the moment, such that 
$\partial_t\hat{\pmb{\Lambda}}^2 = 0$. Now, it is immediate to realize that
$\big[\hat{\pmb{\Lambda}}^2,B(\hat{\mathbf{J}}' -\hat{\pmb{\Lambda}})^2\big] = 0$
since $\hat{\mathbf{J}}'$ and $\hat{\pmb{\Lambda}}^2$ act on different Hilbert spaces (consequently, they
commute with each other) and $[\hat{\Lambda}_i,\hat{\pmb{\Lambda}}^2]=0$. 
One can also 
show that
\begin{equation}
\big[ \hat{\pmb{\Lambda}}^2, \sum_{k\lambda\mu} \omega_B(k) \hat{b}^{\dagger}_{k\lambda\mu}\hat{b}_{k\lambda\mu}\big] = 0\;.
\label{commutator lambda square bosons}
\end{equation}
An explicit derivation of the equation above is provided in App. \ref{app:1}, but 
we also notice that $\hat{\pmb{\Lambda}}^2$ is the Cartan invariant of the 
Lie algebra SO(3) spanned by $\hat{\pmb{\Lambda}}$ as in Eq. \eqref{angular momentum of the bath}.
Since $\hat{H}_B$ is proportional to the identity (in the angular momentum space), 
Eq. \eqref{commutator lambda square bosons} follows immediately.
Therefore, we are left with the task of computing
$\big[\hat{\pmb{\Lambda}}^2,\hat{\mathcal{H}}_C\big]$, where $\hat{\mathcal{H}}_C = \sum_{k\lambda} 
\mathcal{V}_{\lambda}(k)\big(\hat{b}^{\dagger}_{k\lambda 0} + \hat{b}_{k\lambda 0}\big)$ as in 
Eq. \eqref{hamiltonian molecular frame}. As detailed in App. \ref{app:1}, 
the final result is given by 
\begin{equation}
\begin{aligned}
\big[\hat{\pmb{\Lambda}}^2,\hat{\mathcal{H}}_C\big] & = 
\sum_{l = \pm 1,0}\sum_{k\lambda\mu}
(-1)^l \mathcal{V}_{\lambda}(k) 
\bigg[ 2 (\sigma^{\lambda}_{\mu0})_{-l}\;\hat{b}^{\dagger}_{k\lambda\mu}\hat{\Lambda}_{l}\\
& \qquad \qquad + \lambda(\lambda +1) \hat{b}^{\dagger}_{k\lambda 0} 
\delta_{\mu 0}
\delta_{l,\pm 1} - \text{h.c}
\bigg]\;.
\end{aligned}
\label{commutator lambda square coupling more compact}
\end{equation}
Let us also remark that, when considering the hermitian conjugate, 
$\hat{\Lambda}^{\dagger}_l = \hat{\Lambda}_l$ (cfr. the definition in 
Eq. \eqref{angular momentum of the bath}).
For instance, the first term in the
equation above reads, after hermitian conjugation, 
$(\sigma^{\lambda}_{0\mu})_{-l}\;\hat{\Lambda}_l \;\hat{b}_{k\lambda\mu}$.

In order to devise a convenient scheme to compute 
$\langle [\hat{\pmb{\Lambda}}^2,\hat{\mathcal{H}}_C] \rangle$, we 
consider a second unitary transformation 
(see Ref. \onlinecite{devreese-review} for a similar procedure on the Fr\"ohlich Hamiltonian), 
specified by the following operator
\begin{equation}
\hat{U} = \exp\bigg\lbrace 
\sum_{k\lambda}\frac{\mathcal{V}_{\lambda}(k)}{\omega_B(k)}\big( 
\hat{b}_{k\lambda 0} - \hat{b}^{\dagger}_{k\lambda 0}
\big)
\bigg\rbrace\;,
\label{unitary operator slow impurity}
\end{equation}
such that the ground state of the transformed Hamiltonian now corresponds
to the vacuum of environmental excitations.
Therefore, similarly to Ref. \onlinecite{lemeshko-2016}, we can
introduce a variational one-phonon 
ansatz (in the transformed frame given by $\hat{S}$ and $\hat{U}$ as in Eqs. \eqref{unitary transformation matrix angles} and \eqref{unitary operator slow impurity}), namely
\begin{equation}
\ket{\psi_{LM}} = g_{LM}\ket{0} \ket{LM0} + 
\sum_{k\lambda n} \alpha_{k\lambda n} \hat{b}^{\dagger}_{k\lambda n}\ket{0}\ket{LMn}\;.
\label{one-phonon ansatz}
\end{equation}
Here, it is crucial to remark that the above ansatz looks particularly transparent only in the 
transformed frame, i.e. after performing the unitary transformation $\hat{S}$ and $\hat{U}$ defined
in Eqs. \eqref{unitary transformation matrix angles} and \eqref{unitary operator slow impurity} upon the original 
angulon Hamiltonian $\hat{H}_A$ in Eq. \eqref{angulon hamiltonian lab frame}. 
Moving back to the laboratory frame, the state defined in Eq. \eqref{one-phonon ansatz} is 
actually extremely involved, since it entangles the rotor with the infinite degrees of freedom provided by 
the bath.  
Technical details concerning the calculation of $\braket{[\hat{\pmb{\Lambda}}^2,\hat{\mathcal{H}}_C]}$
are extensively provided in App. \ref{app:2}. Here, we just report the final (and remarkably 
simple) result, namely 
\begin{equation}
\bigg\langle \frac{d\hat{\pmb{\Lambda}}^2}{dt}\bigg\rangle = -4 \sum_{k\lambda} 
\lambda(\lambda +1)\;
\mathcal{V}_{\lambda}(k) \;\text{Im}\big(g_{LM}\alpha^*_{k\lambda 0}\big) \;.
\label{final result after average}
\end{equation}
We remark here that the trial wavefunction $\ket{\Psi}$ as in
Eq. \eqref{one-phonon ansatz} (analogous to the Chevy ansatz for the standard polaron framework), 
can be extended to dynamical scenarios by allowing a time dependence
of the complex variational parameters $g_{LM}$ and $\alpha_{k\lambda n}$.
This is achieved by considering, for instance, the time-dependent Lagrangian functional
$\mathcal{L} = \bra{\Psi}\big( i\partial_t - \hat{\mathrm{H}}\big)\ket{\Psi}$ 
\footnote{for details about a more careful formulation for variational approaches, 
see Refs. \cite{kramer-book,frenkel-book}. Conditions for which the main formulations agree 
with each other are usually easy to satisfy, with some caveats analyzed in 
\cite{broeckhove-1988,martinazzo-2019}.}.
\begin{figure}
    \centering
    \begin{subfigure}
    \centering
    \includegraphics[width=\columnwidth]{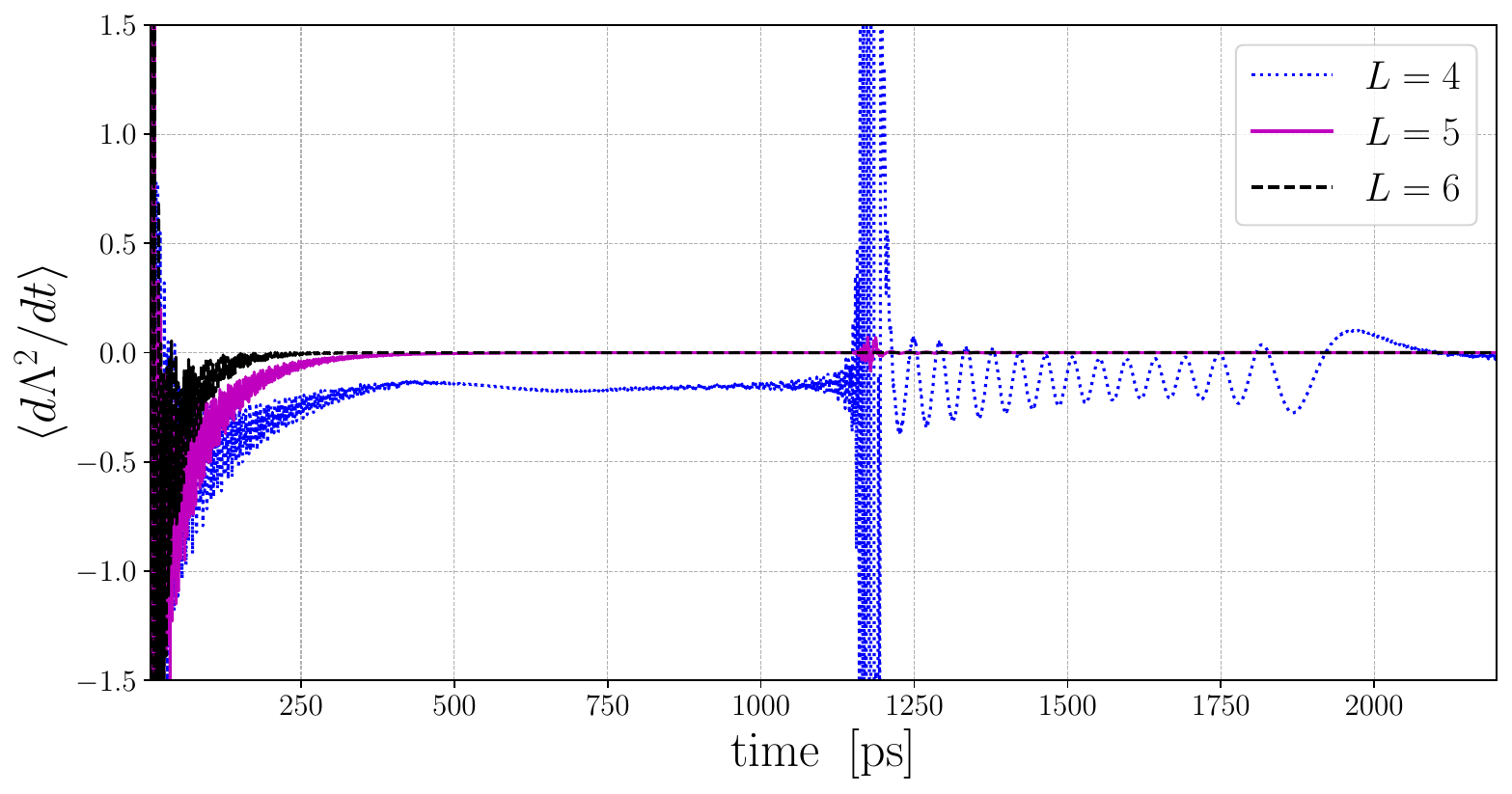}
    \end{subfigure}
    \caption{Time evolution of $\langle d\hat{\pmb{\Lambda}}^2 / dt\rangle$ 
    $L = 4$, $5$ and $L = 6$. All the relevant parameters are as in
    Fig. \ref{fig:2}, but the numerical run has been carried on for longer times. 
    Long-term dynamics confirms that the channel for angular momentum exchange
    is definitely closed for $L > 5$, while a more complex dynamics is observed for 
    lower values of angular momentum, with the onset of intense and fast revivals for 
    $t \gtrsim 10^3$ ps, followed by a low-frequency relaxation around $t \simeq 2\cdot 10^3$ ps.
    In the panel above, we have reported the case of $L = 4$, but a similar behaviour is displayed
    by $L = 2$ and $L = 3$. We identified $L = 5$ as a threshold value, since we can still observe
    a very weak revival while, at the same time, the quasiparticle weight damps to zero 
    (cfr. Fig. \ref{fig:4}).
    }
    \label{fig:3}
\end{figure}
In our case, $\hat{\mathrm{H}} = \hat{U}^{-1}\hat{\mathcal{H}}_{\text{mol}}\hat{U}$, with
$\hat{U}$ as in Eq. \eqref{unitary operator slow impurity} and $\hat{\mathcal{H}}_{\text{mol}}$
being defined in Eq. \eqref{hamiltonian molecular frame}. The variational parameters, 
as mentioned above, are $g_{LM}(t)$ and $\alpha_{k\lambda n}(t)$, which are encoded in 
the trial wavefunction $\ket{\Psi}$ given by Eq. \eqref{one-phonon ansatz}. 
Therefore, from $\mathcal{L}[g_{LM}(t),\alpha_{k\lambda n}(t)]$, time evolution for the 
variational parameters is obtained, as usual, by solving the Euler-Lagrange equations, i.e.
$\frac{d}{dt}\partial_{\dot{x}_i}\mathcal{L} - \partial_{x_i}\mathcal{L} = 0$. 
More details about the lengthy but albeit standard derivation are provided in 
App. \ref{app:3}. 
Now, looking at the impurity-environment coupling in Eq. \eqref{angulon hamiltonian lab frame}, all rotational states are coupled with each other, with Clebsch-Gordan coefficients 
enforcing conservation of angular momentum for each $\lambda$-channel. On the other hand, calculations
are significantly simplified by acknowledging that only few terms are actually relevant 
\cite{szalewicz-2008,stone-book,Farrokhpour-2013,Zang-2014,lemeshko-2017-prl}; for instance, 
in homonuclear molecules such as $I_2$, the quadrupole contribution 
$\lambda = 2$ is the dominant one.
In Fig. \ref{fig:1} we report the time evolution of the quantity 
$\langle d\hat{\pmb{\Lambda}}^2/dt\rangle$ as defined by Eq. \eqref{final result after average}
for a prototypical homonuclear diatomic molecule. 
For those channels, in dimensionless units of
$B$ (energy) and $(mB)^{-1/2}$ (length), we have $\tilde{u}_0 = 218$, $\tilde{u}_2 = 320$ and 
$\tilde{r}_0 = \tilde{r}_2 = 1.5$.
The variational parameters appearing in Eq. \eqref{one-phonon ansatz}
are derived by solving numerically the set of differential equations detailed in App. \ref{app:3}, 
corresponding to the stationary trajectory of the Lagrangian discussed above.  

The physical parameters discussed above are chosen in such a way to reproduce
the characteristic magnitude and range for a homonuclear diatomic molecule in a Helium environment such
as, for instance, $I_2$. Within the effective framework provided by the angulon theory outlined above,
we can observe the following dynamical evolution, displayed in Fig. \ref{fig:2}: 
first, there is an initial transient regime, common to all values of $L$ considered here,
in which an oscillatory decay is observed.
After that, we distinguish two scenarios, depending on the value of $L$. 
For $L > 5$, the decay of $\braket{d\hat{\pmb{\Lambda}^2/dt}}$ continues until
this channel for angular momentum transfer closes completely. This oscillatory behaviour
highlights the exchange of angular momentum between the impurity and the environment; 
a clear example is provided by the $L = 9$ panel in Fig. \ref{fig:2}, where the oscillations
around zero point to a back-and-forth exchange between the rotor and the bosonic bath. For 
$L = 6$, $\braket{d\hat{\pmb{\Lambda}}^2/dt}$ maintains itself negative for the most part 
of the relaxation interval, signalling a gradual slowing down of angular momentum transfer, 
until it gets completely frozen on a similar timescale of the $L = 9$ case. 
On the other hand, a more complex relaxation dynamics of $\braket{d\hat{\pmb{\Lambda}}^2/dt}$
is observed at lower values of $L$, with the $L = 3$ and $L=4$ being reported in Fig. \ref{fig:2}.
Indeed, $\braket{d\hat{\pmb{\Lambda}^2/dt}}$ seems to converge at a negative value, where
the environment is actually leaking angular momentum and transferring it back to the rotor
at a constant rate. However, longer numerical runs (cfr. Fig. \ref{fig:3}) show the onset of revivals
at a similar timescale for $L \lesssim 5$, followed by low-frequency oscillations (if compared
to the initial one). Finally, the channel closes at $t \simeq 2\cdot 10^3$ ps, a 
much longer timescale than the one observed for $L > 5$. 
\begin{figure}[ht!]
\centering
\begin{subfigure}
\centering
\includegraphics[width=\columnwidth]{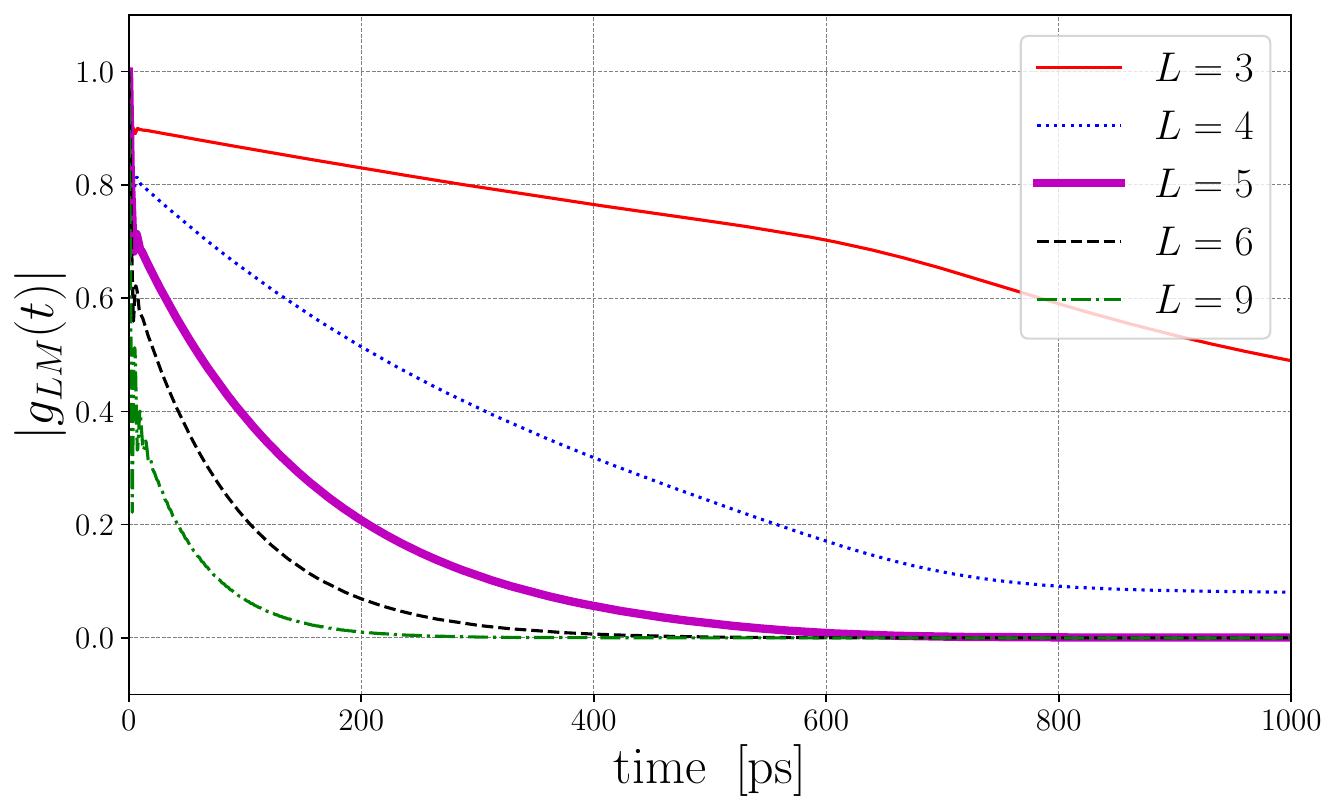}
\end{subfigure}
\caption{Time evolution of $|g_{LM}(t)|$, computed for I$_2$, at different
values of $L$. The complex time-dependent variational parameter $g_{LM}(t)$ 
is introduced in Eq. \eqref{one-phonon ansatz}. Concerning the numerical procedure,
all the relevant parameters are the same as in Fig. \ref{fig:1}. 
Starting from $L = 5$, $g_{LM}$ relaxes to zero, at 
increasingly shorter timescales for higher values of $L$. 
On the contrary, this is not the case when $L \lesssim 5$: this is evident for $L = 3$
(solid red line) and, more remarkably, in the case $L = 4$ we observe $g_{LM}$
relaxing to a non-zero value, which is sustained for a very long time (beyond what is shown
in the panel above).
}
\label{fig:4}
\end{figure}

In Fig. \ref{fig:4} we provide additional insight on this exchange dynamics and the reason for
treating $L = 5$ as a threshold value. There, we report the time evolution of the 
variational parameter $g_{LM}(t)$ for different preparation of the initial state, i.e. 
different values of $L$. Here, we recall that, since the ansatz given by Eq. \eqref{one-phonon ansatz}
describes a linear rotor dressed by bosonic excitations, $g_{LM}(t)$ can be interpreted as the
corresponding quasiparticle weight, pointing to the overlap between the dressed state and the 
bare one $|0\rangle |LM\rangle$, with $|LM\rangle$ being
the eigenstate of an isolated linear rotor. At $t = 0$, the impurity is ideally prepared
in a molecular state distinguished by the desired value of $L$, such that $g_{LM} = 1$. After that,
dynamical evolution unfolds as in Fig. \ref{fig:4}. Initially, $g_{LM}(t)$ displays a non-monotonic
behaviour for all $L$. This initial phase is immediately followed by a relatively fast decay
to zero for $L > 5$, the cases $L = 6$ and $L = 9$ being displayed in Fig. \ref{fig:4}.    
At lower values of $L$, the decay is much slower, with $g_{30}\simeq 0.8$ at $t = 500$ ps.
More remarkably, we see that, with an initial state prepared at $L = 4$ (dotted blue line in 
Fig. \ref{fig:4}), it does not relax to zero and instead
converges to residual quasiparticle weight $\sim 0.1$, surviving for the whole numerical run.
Therefore, when the initial state is 
prepared at low values of $L$, transfer of angular momentum between the rotor and environment
is actually driven by a dressing mechanism, with the impurity effectively behaving as a 
\textit{renormalized} rotor. 
A similar scenario is observed for a point-like impurity moving in a conventional superfluid
(i.e. a U(1) is spontaneously broken, as in Bose-Einstein condensates made of alkali atoms),
where a similar dressing procedure involving Goldstone modes is responsible for transfer of
linear momentum and the onset of a drag force 
\cite{brand-2011,roberts-2006}. This is observed for 
$L \lesssim 5$; we recall here that physical parameters have been chosen to match the 
characteristic magnitude and range of a homonuclear diatomic molecule in a liquid Helium bath.
Therefore, according to our framework, Fig. \ref{fig:4} also clarifies why the $L = 5$ case is 
peculiar. Indeed, at this value of the initial angular momentum, the quasiparticle weigth 
$g_{LM}(t)$ damps to zero for the first time.
This signals that 
the impurity is \textit{effectively losing} its quasiparticle character. Jointly, the 
angular momentum transfer channel driven by the dressing mechanism is closed, since 
phonons appear not to follow the motion of the impurity. 
This dynamical picture is reminiscent  of the angulon instability explored in \cite{lemeshko-2016},
where, by increasing the impurity rotational constant 
(here $B$ in Eq. \eqref{angulon hamiltonian lab frame}),
it becomes unfavourable for excitations to follow rotor's motion and, consequently, 
the bosonic environment does not possess a finite value of angular momentum. 
Interestingly, we can draw a significant parallel with the case of the iodine
molecule in a Helium nanodroplet in the presence of a non-resonant laser 
driving\cite{shepperson-2017}. There, it has been shown that coherent rotations for I$_2$
strongly depend on the laser pulse fluence, i.e. on how that state is prepared initially. 
For low fluences, the molecule and its solvation shell rotates coherently and revivals
are observed 
(in the case of Ref. \onlinecite{shepperson-2017} for the alignment cosine 
$\braket{\cos^2\theta_{2d}}$). On the other hand, for higher fluences the molecule 
\textit{breaks free} from its solvation shell and its dynamics resembles the one
in the gas phase. 

The above discussion well connects to the issue of experimental implementations
and relevance for our theoretical findings. Indeed, from Fig. \ref{fig:4} we realize how
the two different regimes we describe above in Fig. \ref{fig:2} and Fig. \ref{fig:3} 
are due to the dynamical behaviour of the
quasiparticle weight, cfr. Fig. \ref{fig:4}.
As mentioned above, $g_{LM}(t)$ technically quantifies the overlap between the isolated
rotor state and the one dressed by the environment's excitations. Now, such overlap can 
be imaged by means of Ramsey interferometry or spin-echo techniques 
\cite{goold-2011,zangara-2012,knap-2012,schmidt-2016}. More precisely, once one has an overlap 
function generally defined as $\mathcal{S}(t) = \langle \Psi_B |e^{i\hat{H}_0\,t} e^{-i\hat{H}_{\mathrm{tot}}t}| \Psi_B \rangle$, 
with $\hat{H}_0$ ($\hat{H}_{\mathrm{tot}}$) the Hamiltonian in absence of (including) the impurity 
and $\Psi_B$ the initial state of the environment, it is possible to access that 
absorption spectrum $A(\omega) \propto \mathrm{Re}\int_0^{\infty} e^{i\omega t} \mathcal{S}(t)$,
a method pioneered in the context of large impurities embedded in ultracold atomic gases.
In addition, we also remark that a vanishing quasiparticle weight implies the emergence 
of an incoherent background of excitations, thus inhibiting the onset of
rotational revivals (cfr. Fig. \ref{fig:3}), a picture consistent with the 
experimental observations about the iodine molecule made in Ref. \onlinecite{shepperson-2017}.
Concerning the amount of angular momentum exchanged between the rotor and the bosonic medium, 
one can exploit the fact that the superfluid fraction and the average angular momentum of the
environment are intimately connected. Indeed, as in Leggett's 
definition\cite{leggett-1970,leggett-1998}
$\rho_s/\rho =1 - \lim_{\omega\rightarrow 0} (\braket{\pmb{\Lambda}}/I_{\mathrm{cl}}\omega)$,
with $I_{\mathrm{cl}}$ the classical moment of inertia (proportional to $1/B$, $B$ as in 
Eq. \eqref{angulon hamiltonian lab frame}) and $\omega$ the angular velocity of the 
bosonic sample. In ultracold atomic ensembles, the latter can be controlled by 
means of Laguerre-Gaussian beams with different orbital momentum, which also couple 
hyperfine states of the atomic species composing the environments (for instance, 
for $^{87}$Rb, the hyperfine states $m = 0, \pm 1$ in the $F= 1$ manifold). Spectroscopic 
techniques\cite{cooper-2010} can then be used to extract the average occupation number 
in these states. Ideally, the same method can be employed to image changes in the superfluid
fraction related to the presence of the rotating impurity.
%
\section{Conclusions and perspectives}
In this paper we have explored how angular momentum is exchanged between a rotating impurity 
(here modelled as a linear rotor) and
a bosonic environment in its ordered phase (e.g. $^4$He below the $\lambda$-point or an atomic BECs).
We have shown that, depending on the initial preparation of the molecular state (i.e. the value of $L$),
transfer of angular momentum can be driven by the onset of the angulon quasiparticle ($L \lesssim 5$), where
the impurity is effectively dressed by medium fluctuations and rotational dynamics is 
significantly modified. 
On the other hand, for higher values of $L$, this channel is closed due to the fact that the rotor seems
to be highly perturbed by the medium fluctuations, 
resulting in a relatively fast decay of the quasiparticle weight
(see Fig. \ref{fig:4}) and a back-and-forth exchange of angular momentum between the impurity 
and the environment until $\braket{d\hat{\pmb{\Lambda}}^2/dt} \rightarrow 0$, 
as reported in Fig. \ref{fig:2}. 

In the end, this work represents a first step 
towards a more comprehensive understanding  
of dynamical processes involving composite impurities (as, for instance, diatomic molecules)
in many-body environments. 
Even by restricting
to the case of bosonic environment, the picture ahead looks promisingly rich: 
for instance, one could consider 
the presence of an external driving protocol with an explicit time-dependence. 
A similar scenario is considered in 
Refs. \onlinecite{cherepanov-2021,Cherepanov-2022} in relation to alignment dynamics of small molecules in Helium nanodroplets. 
Similarly, at the interesection between chemical physics and quantum optics,
driven-dissipative impurity models involving molecules in cavities are the subject of an intense research effort. 
There, quasiparticles with polariton character may help enhancing the efficiency of certain chemical reactions
\cite{spano-2016,ribeiro-2018,nori-2023}, 
not to mention the inviting perspective of engineering molecular qubits 
\cite{wasielewski-2020} or the possibility to probe predictions from quantum thermodynamics 
\cite{Stickler-2021,gaida2024otto}.   
Moreover, besides more \textit{conventional} superfluids, the interplay between rotational dynamics
and many-body environments is extremely relevant for a variety of scenario, such as Einstein-de Haas phenomenology
in condensed matter and spintronics devices \cite{katsnelson-2019,izumida-2022} or quantum transport in
carbon nanotubes \cite{ambrosetti-2023}.

\begin{acknowledgments}
We thank Henrik Stapelfeldt  for enlightning discussions. 
M.L. acknowledges support by the European Research
Council (ERC) Starting Grant No.801770 (ANGULON).
A.~C. received funding from the European Union’s
Horizon Europe research and innovation program under the
Marie Skłodowska-Curie grant agreement No. 101062862 - NeqMolRot.
\end{acknowledgments}

\section*{Data Availability Statement}
The code for solving Eqs. \eqref{equation of motion for alpha k lambda general} is available
upon reasonable request. 


\onecolumngrid
\appendix

\section{Technical details related to $\langle d\hat{\pmb{\Lambda}}^2/dt\rangle$ - Commutators}
\label{app:1}

\subsection{Derivation of Eq. \eqref{commutator lambda square bosons}}

Here, we aim to show algebraically (cf. Eq. \eqref{commutator lambda square bosons} in the main text) that
\begin{equation}
[\hat{\pmb{\Lambda}}^2,\sum \omega_B(k)
\hat{b}_{k\lambda\mu}^{\dagger}\hat{b}_{k\lambda\mu}] = 0\;.
\label{commutator bath lambda square appendix 1}
\end{equation}
A much shorter group theoretical argument is provided in the main text.
In order to fulfill this task, we first notice that, in terms of spherical components 
\cite{varshalovitch-book},
\begin{equation}
\hat{\pmb{\Lambda}^2} = \sum_{l=\pm 1,0} (-1)^l \hat{\Lambda}_{-l} \hat{\Lambda}_l\;,
\label{norm in spherical components}
\end{equation}
where (cfr. Eq. \eqref{angular momentum of the bath} in the main text) 
$\hat{\Lambda}_l = \sum_{k\lambda\mu\nu} \hat{b}_{k\lambda\mu}^{\dagger}
(\sigma^{\lambda}_{\mu\nu})_l \hat{b}_{k\lambda\nu}$ and 
\begin{equation}
\begin{aligned}
(\sigma^{\lambda}_{\mu\nu})_{\pm 1} & = \sqrt{\lambda(\lambda +1) -\mu(\mu\pm 1)}\, \delta_{\nu,\mu\pm 1} \\
(\sigma^{\lambda}_{\mu\nu})_0 & = \mu\, \delta_{\mu \nu}\;.
\end{aligned}
\label{definition of sigma matrices}
\end{equation}
Therefore, one has to tackle down the following object
\begin{equation}
[\hat{\pmb{\Lambda}}^2,\sum \omega_B(k)
\hat{b}_{k\lambda\mu}^{\dagger}\hat{b}_{k\lambda\mu}] = 
\sum_l \sum_{k\lambda\mu}\sum_{'}\sum_{''}
(-1)^l\omega_B(k) (\sigma^{\lambda'}_{\mu'\nu'})_{-l} (\sigma^{\lambda''}_{\mu''\nu''})_{l}
\bigg[ 
\hat{b}^{\dagger}_{k'\lambda'\mu'}\hat{b}_{k'\lambda'\nu'}\hat{b}^{\dagger}_{k''\lambda''\mu''}\hat{b}_{k''\lambda''\nu''},
\hat{b}^{\dagger}_{k\lambda\mu}\hat{b}_{k\lambda\mu}
\bigg]
\label{to compute long}
\end{equation}
where $\sum_{'} = \sum_{k'\lambda'\mu'\nu'}$ and analogously for $\sum_{''}$. Now, for the sake of clarity,
let us rename
\begin{equation}
\begin{aligned}
 A & \longrightarrow k'\lambda'\mu' \qquad A' \longrightarrow k''\lambda''\mu''\\
 B & \longrightarrow k'\lambda'\nu' \qquad B' \longrightarrow k''\lambda''\nu''\\
 C & \longrightarrow k\lambda\mu \;.
\end{aligned}
\label{renaming}
\end{equation}
Therefore, the commutator $[\hat{b}^{\dagger}_{A}\;\hat{b}_{B}\;\hat{b}^{\dagger}_{A'}\;\hat{b}_{B'},\;\hat{b}^{\dagger}_C\;\hat{b}_C]$ can be simplified by reducing both terms to their normal form
(creation operators on the left, annihilation ones on the right), such that
\begin{equation}
\begin{aligned}
[\hat{b}^{\dagger}_{A}\;\hat{b}_{B}\;\hat{b}^{\dagger}_{A'}\;\hat{b}_{B'},\;\hat{b}^{\dagger}_C\;\hat{b}_C] & 
=  \hat{b}^{\dagger}_{A}\;\hat{b}^{\dagger}_{A'}\;\hat{b}_{B}\;\hat{b}_{C} \delta_{B'C} 
 - \hat{b}^{\dagger}_{C}\;\hat{b}^{\dagger}_{A}\;\hat{b}_{B}\;\hat{b}_{B'} \delta_{A'C} \\ 
& \qquad \qquad + \hat{b}^{\dagger}_{A}\;\hat{b}^{\dagger}_{A'}\;\hat{b}_{B'}\;\hat{b}_{C} \delta_{BC} 
-\hat{b}^{\dagger}_{C}\;\hat{b}^{\dagger}_{A'}\;\hat{b}_{B}\;\hat{b}_{B'} \delta_{AC} \\
 & \qquad \qquad \qquad \qquad +  \hat{b}^{\dagger}_{A}\;\hat{b}_{C} \delta_{A'B}\delta_{B'C} 
 - \hat{b}^{\dagger}_{C}\;\hat{b}_{B} \delta_{AC}\delta_{A'B} \;.
\end{aligned}
\label{the commutator}
\end{equation}
It can be shown that every term in the equation above provides a null contribution. 
For instance, the last line is evidently zero once we let $\delta_{B'C}$ and $\delta_{AC}$ act on the
respective terms. The other two terms require some additional work:
if we take into account the first line on the RHS of Eq. \eqref{the commutator} and 
the summations in Eq. \eqref{to compute long}, it reads
\begin{equation}
\begin{aligned}
(\bullet)& =\sum_{k\lambda\mu}\sum_{'}\sum_{''} 
(\sigma^{\lambda'}_{\mu'\nu'})_{-l} (\sigma^{\lambda''}_{\mu''\nu''})_{l}
\bigg[ 
\hat{b}^{\dagger}_{k'\lambda'\mu'} \hat{b}^{\dagger}_{k''\lambda''\mu''} \hat{b}_{k'\lambda'\nu'} \hat{b}_{k\lambda\mu} \;\delta_{kk''}\delta_{\lambda\lambda''}\delta_{\mu\nu''} 
- \hat{b}^{\dagger}_{k\lambda\mu} \hat{b}^{\dagger}_{k'\lambda'\mu'} \hat{b}_{k''\lambda'\nu''} 
\hat{b}_{k'\lambda'\nu'} \; \delta_{kk''}\delta_{\lambda\lambda''}\delta_{\mu\mu''} 
\bigg] =\\
&= \sum_{k\lambda\mu}\sum_{'}(\sigma^{\lambda'}_{\mu'\nu'})_{-l}\Bigg[
\sum_{\mu''}(\sigma^{\lambda}_{\mu''\mu})_l \hat{b}^{\dagger}_{k'\lambda'\mu'} \hat{b}^{\dagger}_{k\lambda\mu''} \hat{b}_{k'\lambda'\nu'} \hat{b}_{k\lambda\mu}
- \sum_{\nu''}(\sigma^{\lambda}_{\mu\nu''})_l \hat{b}^{\dagger}_{k\lambda\mu} \hat{b}^{\dagger}_{k'\lambda'\mu'} \hat{b}_{k\lambda\nu''} \hat{b}_{k'\lambda'\nu'}
\bigg] \\
& = \sum_{k\lambda}\sum_{k'\lambda'\mu'\nu'}(\sigma^{\lambda'}_{\mu'\nu'})_{-l}
\bigg[ 
\sum_{\mu\mu''} (\sigma^{\lambda}_{\mu''\mu})_l  \hat{b}^{\dagger}_{k'\lambda'\mu'} \hat{b}^{\dagger}_{k\lambda\mu''} \hat{b}_{k'\lambda'\nu'} \hat{b}_{k\lambda\mu} -
\underbrace{\sum_{\mu\nu''}  (\sigma^{\lambda}_{\mu\nu''})_l \hat{b}^{\dagger}_{k\lambda\mu} \hat{b}^{\dagger}_{k'\lambda'\mu'} \hat{b}_{k\lambda\nu''} \hat{b}_{k'\lambda'\nu'}}_{\text{rename $\mu\rightarrow\mu''$ and $\nu''\rightarrow \mu$}}
\bigg] = 0\;.
\end{aligned}
\label{first line to show it is zero}
\end{equation}
It is important to remark that the algebraic manipulations described above holds under the 
relevant assumption the bosonic spectrum of the environment does not depend on the orbital or 
magnetic numbers. Then, the exactly same argument can be applied to the second line in Eq. \eqref{the commutator}, such that
Eq. \eqref{commutator lambda square bosons} in the main text is now proved. 
\subsection{Derivation of Eq. \eqref{commutator lambda square coupling more compact}}

The starting point is obviously given by 
\begin{equation}
\bigg[ \hat{\pmb{\Lambda}}^2, \sum_{k\lambda} \mathcal{V}_{\lambda}(k)\big(
\hat{b}^{\dagger}_{k\lambda 0} + \hat{b}_{k\lambda 0} 
\big)\bigg] 
= \sum_{k\lambda}\mathcal{V}_{\lambda}(k) \bigg( 
\big[\hat{\pmb{\Lambda}}^2,\hat{b}^{\dagger}_{k\lambda 0}\big] + 
\big[\hat{\pmb{\Lambda}}^2,\hat{b}_{k\lambda 0} \big]
\bigg)\;.
\label{commutator coupling step 1}
\end{equation}
By making use of Eq. \eqref{norm in spherical components} and Eq. \eqref{angular momentum of the bath}
in the main text, one gets, for instance,
\begin{equation}
\big[\hat{\pmb{\Lambda}}^2,\hat{b}^{\dagger}_{k\lambda 0}\big] 
= \sum_{l = \pm 1, 0}(-1)^{l} \sum_{'}\sum_{''} (\sigma^{\lambda'}_{\mu'\nu'})_{-l} (\sigma^{\lambda''}_{\mu''\nu''})_{l}
\bigg[ 
\hat{b}^{\dagger}_{k'\lambda'\mu'}\hat{b}_{k'\lambda'\nu'}\hat{b}^{\dagger}_{k''\lambda''\mu''}\hat{b}_{k''\lambda''\nu''},\hat{b}^{\dagger}_{k\lambda 0} \bigg]
\label{commutator coupling step 2}
\end{equation}
where, similarly to the previous section, 
$\sum_{'} = \sum_{k'\lambda'\mu'\nu'}$ and analogously for $\sum_{''}$. 
In the same fashion of Eq. \eqref{renaming}, we define
\begin{equation}
\begin{aligned}
 A & \longrightarrow k'\lambda'\mu' \qquad A' \longrightarrow k''\lambda''\mu''\\
 B & \longrightarrow k'\lambda'\nu' \qquad B' \longrightarrow k''\lambda''\nu''\\
 C & \longrightarrow k\lambda 0 \;,
\end{aligned}
\label{renaming}
\end{equation}
such that
\begin{equation}
\big[\hat{b}^{\dagger}_A\;\hat{b}_B\;\hat{b}^{\dagger}_{A'}\;\hat{b}_{B'}, \hat{b}^{\dagger}_C\big] =
\hat{b}^{\dagger}_A\;\hat{b}^{\dagger}_{A'}\;\hat{b}_{B'}\;\delta_{BC} +
\hat{b}^{\dagger}_A \;\hat{b}^{\dagger}_{A'}\;\hat{b}_B \;\delta_{B'C} +
\hat{b}^{\dagger}_{A}\;\delta_{A'B}\;\delta_{B'C}\;.
\label{commutator coupling with a b c}
\end{equation}
Now, putting back the physical indices, one recovers the following equation
\begin{equation}
\begin{aligned}
\big[\hat{\pmb{\Lambda}}^2,\hat{b}^{\dagger}_{k\lambda 0}\big] & =
\sum_{l =\pm 1,0}(-1)^l\sum_{'}\sum_{''} (\sigma^{\lambda'}_{\mu'\nu'})_{-l} (\sigma^{\lambda''}_{\mu''\nu''})_{l} \bigg[
\hat{b}^{\dagger}_{k'\lambda'\mu'} \;\hat{b}^{\dagger}_{k''\lambda''\mu''}\;\hat{b}_{k''\lambda''\nu''}\;\delta_{kk'}\delta_{\lambda\lambda'}\delta_{\nu' 0} \\
& \qquad\qquad +\hat{b}^{\dagger}_{k'\lambda'\mu'} \;\hat{b}^{\dagger}_{k''\lambda''\mu''}\;\hat{b}_{k'\lambda'\nu'}\;\delta_{kk''}\delta_{\lambda\lambda''}\delta_{\nu'' 0} + 
\hat{b}^{\dagger}_{k'\lambda'\mu'}\;
\delta_{k''k}\delta_{k''k}\;\delta_{\lambda''\lambda'}\delta_{\lambda''\lambda}\;\delta_{\mu''\nu'}\delta_{\nu''0}
\bigg] \\
& = \sum_{l=\pm 1,0}(-1)^l\bigg[ 
\sum_{\substack{k''\lambda''\mu''\\\nu''\mu'}}(\sigma^{\lambda}_{\mu'0})_{-l} \;(\sigma^{\lambda''}_{\mu''\nu''})_{l} \;\hat{b}^{\dagger}_{k\lambda\mu'}\;\hat{b}^{\dagger}_{k''\lambda''\mu''}\;\hat{b}_{k''\lambda''\nu''} \\
& \qquad \qquad +\sum_{\substack{k'\lambda'\mu'\\\nu'\mu''}}(\sigma^{\lambda'}_{\mu'\nu'})_{-l}\; (\sigma^{\lambda}_{\mu''0})_{l} \;\hat{b}^{\dagger}_{k'\lambda'\mu'}\;\hat{b}^{\dagger}_{k\lambda\mu''}\;\hat{b}_{k'\lambda'\nu'} \\
& \qquad \qquad  \qquad \qquad +\sum_{\substack{k''\lambda''\mu''\\\nu''\mu'}} 
(\sigma^{\lambda''}_{\mu'\mu''})_{-l}\;(\sigma^{\lambda''})_{\lambda''\nu''}\;
\hat{b}^{\dagger}_{k''\lambda''\mu'}\;\delta_{k''k'}\delta_{\lambda'\lambda}\delta_{\nu''0}
\bigg]\;.
\end{aligned}
\label{commutator coupling step 3}
\end{equation}
While the equation above may look definitely cumbersome at first sight, it is possible to recognize 
$\hat{\Lambda}_l$ and $\hat{\Lambda}_{-l}$ in the first and second line of the last step. Therefore,
if we reinsert also the first summation on $k$, and $\lambda$, we are led to 
\begin{equation}
\begin{aligned}
\sum_{k\lambda}\mathcal{V}_{\lambda}(k)\big[\hat{\pmb{\Lambda}}^2,\hat{b}^{\dagger}_{k\lambda 0}\big] & 
=\sum_{l=\pm 1,0}\sum_{k\lambda}(-1)^l\mathcal{V}_{\lambda}(k)\bigg[ 
\sum_{\mu'}(\sigma^{\lambda}_{\mu'0})_{-l} \hat{b}^{\dagger}_{k\lambda\mu'}\;\hat{\Lambda}_l 
+\sum_{\mu''}(\sigma^{\lambda}_{\mu''0})_l \hat{b}^{\dagger}_{k\lambda\mu''}\;\hat{\Lambda}_{-l}
+\sum_{\mu'\mu''}(\sigma^{\lambda}_{\mu'\mu''})_{-l}(\sigma^{\lambda}_{\mu''0})_l \hat{b}_{k\lambda\mu'}
\bigg]\;.
\end{aligned}
\label{commutator coupling step 4}
\end{equation}
Now, for the sake of clarity, 
we are allowed to rename the indices as $\mu' \rightarrow \mu$ and $\mu''\rightarrow \mu'$. In addition,
%
%
Now, in the equation above, the last term (actually involving a matrix product, with index $\mu''$ being contracted)
can be simplified by making use of Eq. \eqref{definition of sigma matrices}. This implies that
\begin{equation}
\sum_{\mu'} \sum_{\mu''} (\sigma^{\lambda}_{\mu'\mu''})_{-l} \; (\sigma^{\lambda}_{\mu''0})_{+l} =
\lambda \big(\lambda +1 \big) \delta_{l,\pm 1} \delta_{\mu 0}\;.
\label{contracting an index}
\end{equation}
Thus, finally, 
\begin{equation}
\sum_{k\lambda}\mathcal{V}_{\lambda}(k)\big[\hat{\pmb{\Lambda}}^2,\hat{b}^{\dagger}_{k\lambda 0}\big] = 
\sum_{l= \pm 1,0}\sum_{k\lambda\mu} (-1)^l \mathcal{V}_{\lambda}(k)\bigg[
(\sigma^{\lambda}_{\mu 0})_{-l}\; \hat{b}^{\dagger}_{k\lambda\mu}\; \hat{\Lambda}_{l}+
(\sigma^{\lambda}_{\mu 0})_l \;\hat{b}^{\dagger}_{k\lambda\mu}\; \hat{\Lambda}_{-l} + 
\lambda(\lambda +1)\;\hat{b}^{\dagger}_{k\lambda 0} \;\delta_{l,\pm 1}
\bigg]\;.
\label{commutator coupling step 5}
\end{equation}
The exact same argument can be employed to compute $\big[\hat{\pmb{\Lambda}}^2,\hat{b}_{k\lambda 0}\big]$, with the important difference of a global minus in the equivalent of 
Eq. \eqref{commutator coupling with a b c}, i.e. 
\begin{equation}
\big[\hat{b}^{\dagger}_A\;\hat{b}_B\;\hat{b}^{\dagger}_{A'}\;\hat{b}_{B'}, \hat{b}_C\big] =
- \hat{b}^{\dagger}_A\;\hat{b}_{B}\;\hat{b}_{B'}\;\delta_{A'C} 
- \hat{b}^{\dagger}_{A'} \;\hat{b}_{B}\;\hat{b}_{B'} \;\delta_{AC} 
- \hat{b}_{B'}\;\delta_{AC}\;\delta_{BA'}\;.
\label{commutator coupling with a b c hermitian cojugate}
\end{equation}
Retracing the steps leading us to Eq. \eqref{commutator coupling step 5} we then verify
that 
\begin{equation}
\begin{aligned}
\big[\hat{\pmb{\Lambda}}^2,\hat{\mathcal{H}}_C\big] & = \sum_{l = \pm 1,0}\sum_{k\lambda\mu}
(-1)^l \mathcal{V}_{\lambda}(k) 
\bigg[ (\sigma^{\lambda}_{\mu0})_{-l}\;\hat{b}^{\dagger}_{k\lambda\mu}\hat{\Lambda}_{l}\\
& +(\sigma^{\lambda}_{\mu 0})_l \;\hat{b}^{\dagger}_{k\lambda\mu}\hat{\Lambda}_{-l}
+ \lambda(\lambda +1) \hat{b}^{\dagger}_{k\lambda 0} \delta_{\mu 0}\delta_{l,\pm 1} - \text{h.c}
\bigg]\;.
\end{aligned}
\label{commutator lambda square coupling}
\end{equation}
Actually, in order to recover Eq. \eqref{commutator lambda square coupling more compact},
we just need to write down the terms for $l = \pm 1$ and $l = 0$, and then
regroup terms equal to each other.
%
%
%
%
%
\section{The single-phonon ansatz - Computing expectation values}
\label{app:2}

Here, we aim to provide the reader with extensive details on the
calculation of the following expectation values 
(cfr. Eq. \eqref{lambda square ehrenfest}
in the main text),
\begin{equation}
\bigg\langle \frac{d\hat{\pmb{\Lambda}}^2}{dt} \bigg\rangle
= -i \big\langle \big[\hat{\pmb{\Lambda}}^2,\hat{\mathcal{H}}_{\text{mol}}\big]\big \rangle\;,
\label{lambda square ehrenfest supplementary}
\end{equation}
with the commutator $\big[\hat{\pmb{\Lambda}}^2,\hat{\mathcal{H}}_{\text{mol}}\big]$ 
being given by Eq. \eqref{commutator lambda square coupling} 
(or the more compact Eq. \eqref{commutator lambda square coupling more compact}).
Before applying the unitary transformation defined by $\hat{U}$ and the one-phonon ansatz,
respectively Eq. \eqref{unitary operator slow impurity}
and Eq. \eqref{one-phonon ansatz} in the main text, let us recast the equation above
in a slightly more convenient form. In particular,
\begin{equation}
\begin{aligned}
\bigg\langle \frac{d\hat{\pmb{\Lambda}}^2}{dt} \bigg\rangle
& = -i \big\langle \big[\hat{\pmb{\Lambda}}^2,\hat{\mathcal{H}}_{\text{mol}}\big]\big \rangle \\
& = - i \sum_{l = \pm 1, 0}\sum_{k\lambda\mu} (-1)^l \mathcal{V}_{\lambda}(k) \bigg\langle
2\,(\sigma^{\lambda}_{\mu 0})_{-l}\; \hat{b}^{\dagger}_{k\lambda \mu} \, \hat{\Lambda}_l
+ \lambda(\lambda +1)\, \hat{b}^{\dagger}_{k\lambda 0} \delta_{\mu 0}\delta_{l,\pm 1} - \text{h.c}
\bigg\rangle \\
& = - i \sum_{l = \pm 1, 0}\sum_{k\lambda\mu} (-1)^l \mathcal{V}_{\lambda}(k) 
\bigg[ 
2\, (\sigma^{\lambda}_{\mu 0})_{-l}\; 
\big\langle \hat{b}^{\dagger}_{k\lambda \mu} \, \hat{\Lambda}_l \big\rangle
+ \lambda (\lambda+1)\,\delta_{\mu 0}\delta_{l,\pm 1} 
\big\langle\hat{b}^{\dagger}_{k\lambda 0} \big\rangle
- \text{h.c}\;.
\bigg]
\end{aligned}
\label{slightly more convenient form for the expectation value}
\end{equation} 
It is then easy to notice that computing 
$\big\langle \frac{d\hat{\pmb{\Lambda}}^2}{dt}\big\rangle$ actually involves just 
$\big\langle\hat{b}^{\dagger}_{k\lambda\mu} \hat{\Lambda}_l\big\rangle$ and
$\big\langle \hat{b}^{\dagger}_{k\lambda 0}\big\rangle$.
\subsection{Computing $\big\langle \hat{b}^{\dagger}_{k\lambda 0}\big\rangle$}
Let us being with the easiest term, i.e. 
$\big\langle \hat{b}^{\dagger}_{k\lambda 0}\big\rangle$. By applying the unitary
transformation defined by $\hat{U}$ in Eq. \eqref{unitary operator slow impurity}, we 
end up with
\begin{equation}
\begin{aligned}
\big\langle \hat{b}^{\dagger}_{k\lambda 0}\big\rangle &=
\big\langle \Psi \big| \hat{U}^{-1} \hat{b}^{\dagger}_{k\lambda 0}\hat{U} \big|\Psi\big\rangle
= \big\langle \Psi \big| \hat{b}^{\dagger}_{k\lambda 0} \big|\Psi\big\rangle
- \frac{\mathcal{V}_{\lambda}(k)}{\omega_B(k)} ||\Psi ||^2\;,
\end{aligned}
\label{shifting the bosonic operator}
\end{equation}
$\big| \Psi\big\rangle$ being given by Eq. \eqref{one-phonon ansatz}. 
We notice that, when considering the hermitian conjugate, the
shifts driven by the unitary transformation $\hat{U}$ cancel with each other.
Now, we are actually ready to plug the one-phonon ansatz into 
Eq. \eqref{shifting the bosonic operator}, such that
\begin{equation}
\begin{aligned}
\big\langle \Psi \big| \hat{b}^{\dagger}_{k\lambda 0} \big|\Psi\big\rangle & =
\bigg(g^*_{LM}\bra{0} \bra{LM0} + 
\sum_{k'\lambda' n'} \alpha^*_{k'\lambda' n'}\bra{0}\bra{LMn'} \hat{b}_{k'\lambda' n'} \bigg)\;
\hat{b}^{\dagger}_{k\lambda 0}\;
\bigg(g_{LM}\ket{0} \ket{LM0} + 
\sum_{k''\lambda'' n''} \alpha_{k''\lambda'' n''} \hat{b}^{\dagger}_{k''\lambda'' n''}\ket{0}\ket{LMn''} \bigg)\\
& \\
& = |g_{LM}|^2 \bra{LM0} \bra{0} \hat{b}^{\dagger}_{k\lambda 0} \ket{0} \ket{LM0} 
+ g^*_{LM}\sum_{k''\lambda''n ''}   \alpha_{k''\lambda'' n''}\;\bra{LM0} \bra{0}
\hat{b}^{\dagger}_{k''\lambda'' n''} \hat{b}^{\dagger}_{k\lambda 0} \ket{0} \ket{LMn''} \\
& \\
& \qquad\qquad +  g_{LM}\sum_{k'\lambda'n'} \alpha^*_{k'\lambda' n'} \bra{0} \bra{LMn'} \hat{b}_{k'\lambda' n'} \hat{b}^{\dagger}_{k\lambda 0} \ket{0} \ket{LM0} \\
& \\
& \qquad \qquad \qquad + \sum_{k'\lambda' n'}\sum_{k''\lambda'' n''} \alpha^*_{k''\lambda'' n''}
\alpha_{k'\lambda' n'} \bra{0} \bra{LMn'} \hat{b}_{k'\lambda' n'} \hat{b}^{\dagger}_{k\lambda 0} \hat{b}^{\dagger}_{k''\lambda'' n''} \ket{0} \ket{LMn''}\\
& \\
& = g_{LM}\sum_{k'\lambda' n'} \alpha^*_{k'\lambda' n'} \braket{LMn'|LM0} \;\bra{0} \hat{b}_{k'\lambda' n'} \hat{b}^{\dagger}_{k\lambda 0} \ket{0} \\
& \\
& = g_{LM}\sum_{k'\lambda' n'} \alpha^*_{k'\lambda'n'}\;\delta_{n'0}\;\delta_{kk'}\;\delta_{\lambda\lambda'} \\
& \\
& = g_{LM} \alpha^*_{k\lambda 0}\;.
\end{aligned}
\label{computing expectation value for a single bosonic operator}
\end{equation}
A completely analogous calculation leads us to 
$\big\langle \Psi \big | \hat{b}_{k\lambda 0}\big| \Psi\big\rangle = g^*_{LM}\alpha_{k\lambda 0}$.
\subsection{Computing $\big\langle\hat{b}^{\dagger}_{k\lambda\mu} \hat{\Lambda}_l\big\rangle$ and
$\big\langle \hat{\Lambda}_l\; \hat{b}_{k\lambda\mu}\big\rangle$}

We now consider the less trivial $\big\langle\hat{b}^{\dagger}_{k\lambda\mu} \hat{\Lambda}_l\big\rangle$.
First, we explore how the unitary transformation act on the operator product; simple manipulations provides
us with the following intermediate result
\begin{equation}
\begin{aligned}
\big\langle\hat{b}^{\dagger}_{k\lambda\mu} \hat{\Lambda}_l\big\rangle & = 
\bra{\Psi}\hat{U}^{-1}\hat{b}^{\dagger}_{k\lambda\mu}\;\hat{\Lambda}_l\; \hat{U} \ket{\Psi} 
= \bra{\Psi}\hat{U}^{-1}\hat{b}^{\dagger}_{k\lambda\mu}\;\hat{U}\hat{U}^{-1}\;\hat{\Lambda}_l\; \hat{U} \ket{\Psi} \\
& \\
& = \bigg\langle \Psi \bigg|
\bigg(\hat{b}^{\dagger}_{k\lambda\mu}  - \frac{\mathcal{V}_{\lambda}(k)}{\omega_B(k)}\delta_{\mu 0}\bigg) 
\hat{U}^{-1}\;\hat{\Lambda}_l\; \hat{U}
\bigg| \Psi \bigg\rangle \\
& \\
& =\bra{\Psi}\hat{b}^{\dagger}_{k\lambda\mu}\;\hat{U}^{-1}\;\hat{\Lambda}_l\; \hat{U} \ket{\Psi} \;,
\end{aligned}
\label{applying the shift transformation to the product}
\end{equation}
where we noticed that the shift comes with a $\delta_{\mu 0}$, reading a factor $(\sigma^{\lambda}_{00})_{-l}$
(cfr. with the first term in Eq. \eqref{commutator lambda square coupling more compact}), which is identically 
zero. Now, components of the environment angular momentum transform as
\begin{equation}
\begin{aligned} 
\hat{U}^{-1}\;\hat{\Lambda}_l\; \hat{U} & = \sum_{k\lambda \mu \nu} (\sigma^{\lambda}_{\mu \nu})_l
\hat{U}^{-1} \hat{b}^{\dagger}_{k\lambda\mu} \hat{b}_{k\lambda\nu} \hat{U} = 
\sum_{k\lambda \mu \nu} (\sigma^{\lambda}_{\mu \nu})_l
\hat{U}^{-1} \hat{b}^{\dagger}_{k\lambda\mu}\hat{U}\hat{U}^{-1} \hat{b}_{k\lambda\nu} \hat{U}\\
& = \sum_{k\lambda\mu \nu}(\sigma^{\lambda}_{\mu \nu})_l
\bigg(\hat{b}^{\dagger}_{k\lambda\mu}  - \frac{\mathcal{V}_{\lambda}(k)}{\omega_B(k)}\delta_{\mu 0}\bigg)
\bigg(\hat{b}_{k\lambda\nu}  - \frac{\mathcal{V}_{\lambda}(k)}{\omega_B(k)}\delta_{\nu 0}\bigg) \\
& = \sum_{k\lambda\mu \nu}(\sigma^{\lambda}_{\mu \nu})_l
\bigg[ 
\hat{b}^{\dagger}_{k\lambda\mu}\hat{b}_{k\lambda\nu} - \frac{\mathcal{V}_{\lambda}(k)}{\omega_B(k)}
\delta_{\nu 0}\;\hat{b}^{\dagger}_{k\lambda\mu} - \frac{\mathcal{V}_{\lambda}(k)}{\omega_B(k)}\delta_{\mu 0}
\;\hat{b}_{l\lambda\nu}  + \frac{\mathcal{V}^2_{\lambda}(k)}{\omega^2_B(k)}\delta_{\mu 0}\delta_{\nu 0}
\bigg] \\
& = \hat{\Lambda}_l - \sum_{k\lambda\mu \nu} \frac{\mathcal{V}_{\lambda}(k)}{\omega_B(k)}
\bigg[(\sigma^{\lambda}_{\mu \nu})_l \;\hat{b}^{\dagger}_{k\lambda\mu} + (\sigma^{\lambda}_{0\mu})_l 
\;\hat{b}_{k\lambda \mu} \bigg] \;.
\end{aligned}
\label{how lambda transforms under u}
\end{equation}
We realize that, again, a factor $(\sigma^{\lambda}_{00})_l$ appears because of the
double delta function in the last term of the second line. As a consequence, that contribution is zero.
Now, we plug the equation above into Eq. \eqref{applying the shift transformation to the product}, 
finding that
\begin{equation}
\begin{aligned}
\big\langle\hat{b}^{\dagger}_{k\lambda\mu} \hat{\Lambda}_l\big\rangle & = 
\bra{\Psi} \hat{b}^{\dagger}_{k\lambda\mu} \hat{\Lambda}_l \ket{\Psi} 
- \sum_{k'\lambda'\mu'}\frac{\mathcal{V}_{\lambda'}(k')}{\omega_B (k')}(\sigma^{\lambda'}_{\mu' 0})_l 
\bra{\Psi}\hat{b}^{\dagger}_{k\lambda\mu} \hat{b}^{\dagger}_{k'\lambda'\mu'} \ket{\Psi}  
- \sum_{k'\lambda'\mu'}\frac{\mathcal{V}_{\lambda'}(k')}{\omega_B (k')}(\sigma^{\lambda'}_{0 \mu'})_l 
\bra{\Psi}\hat{b}^{\dagger}_{k\lambda\mu} \hat{b}_{k'\lambda'\mu'} \ket{\Psi}\;.
\end{aligned}
\end{equation}
Now, it is known (cfr. Ref. \cite{lemeshko-2016})
that $\bra{\Psi}\hat{b}^{\dagger}_{k\lambda\mu} \hat{b}_{k'\lambda'\mu'} \ket{\Psi}
= \delta_{\mu 0}\delta_{\mu' 0} \; (\ldots)$, therefore the last term in the equation above is zero.
On the other hand, within the one-phonon ansatz framework is likewise immediate to realize that
$\braket{\hat{b}^{\dagger}_{k\lambda\mu} \hat{b}^{\dagger}_{k'\lambda'\mu'}} = 0$. 
Therefore, we are left with the fact that the expectation value 
$\big\langle\hat{b}^{\dagger}_{k\lambda\mu} \hat{\Lambda}_l\big\rangle$ is actually transparent under
the action of the unitary operator $\hat{U}$ as given by Eq. \eqref{unitary operator slow impurity}. 
In other words, we just have to consider $\bra{\Psi} \hat{b}^{\dagger}_{k\lambda\mu} \hat{\Lambda}_l \ket{\Psi}$
with $\ket{\Psi}$ given by Eq. \eqref{one-phonon ansatz}. 

However, things can be simplified even further. In fact, if we compute the expectation value
explicitly, we realize that 
\begin{equation}
\begin{aligned}
\big\langle\hat{b}^{\dagger}_{k\lambda\mu} \hat{\Lambda}_l\big\rangle & = 
\bra{\Psi} \hat{b}^{\dagger}_{k\lambda\mu} \hat{\Lambda}_l \ket{\Psi} \\
& \\
& = \bigg( g^*_{LM}\bra{0}\bra{LM0} + \sum_{k'\lambda' n'} \alpha^*_{k'\lambda' n'} 
 \bra{0}\bra{LMn'}\hat{b}_{k'\lambda' n'}\bigg) \; 
\hat{b}^{\dagger}_{k\lambda\mu} \hat{\Lambda}_l \\
& \qquad \qquad  \times \bigg(g_{LM}\ket{0}\ket{LM0} + \sum_{k''\lambda'' n''} \alpha_{k''\lambda'' n''} 
\hat{b}^{\dagger}_{k''\lambda'' n''} \ket{0}\ket{LMn''}\bigg)	\\
& \\
& = |g_{LM}|^2 \bra{LM0}\bra{0} \hat{b}^{\dagger}_{k\lambda\mu} \hat{\Lambda}_l \ket{0}\ket{LM0} \\
&\qquad \qquad +g_{LM} \sum_{k'\lambda' n'} \alpha^{*}_{k'\lambda' n'} \bra{LMn'}\bra{0}
\hat{b}_{k'\lambda'n'}\;\hat{b}^{\dagger}_{k\lambda\mu}\;\hat{\Lambda}_l\ket{0}\ket{LM0}\\
&\qquad \qquad \qquad+ g^*_{LM} \sum_{k''\lambda'' n''} \alpha_{k''\lambda'' n''} \bra{LM0}\bra{0}
\hat{b}^{\dagger}_{k\lambda\mu}\;\hat{\Lambda}_l \;\hat{b}^{\dagger}_{k''\lambda''n''}
\ket{0}\ket{LMn''} \\
& \qquad \qquad \qquad \qquad+ \sum_{k'\lambda' n'} \sum_{k''\lambda'' n''} \alpha^*_{k'\lambda' n'}
\alpha_{k''\lambda'' n''}\bra{LMn'}\bra{0} \hat{b}_{k'\lambda' n'}\;
\hat{b}^{\dagger}_{k\lambda\mu}\; \hat{\Lambda}_l\; \hat{b}^{\dagger}_{k''\lambda'' n''}
\ket{0}\ket{LMn''} \;.
\end{aligned}
\label{first expectation value with lambda}
\end{equation}
Now, $\hat{\Lambda}_l$ acts on a phonon states as \cite{bernath-book}
\begin{equation}
\hat{\Lambda}_l \ket{\lambda,\mu,\nu} = \sqrt{\lambda(\lambda +1)}\, 
C^{\lambda \mu+l}_{\lambda \mu, 1l}\ket{\lambda,\mu+l,\nu}\;.
\label{action of Lambda on a phonon ket}
\end{equation}
Therefore, $\hat{\Lambda}_l\ket{0} = 0$ (since $\lambda = 0$) and, consequently, the first
and the second term in the equation above are identically zero. It is likewise immediate to
realize that, since $\hat{\Lambda}_l$ does not change the number of phonons, the remaining
terms are zero as well, since they involve of a different number of creation and
annihilation operators.
 
Now, by repeating the exact same steps as in 
Eqs. \eqref{applying the shift transformation to the product} and
\eqref{how lambda transforms under u}, we easily get 
\begin{equation}
\big\langle \hat{\Lambda}_l\; \hat{b}_{k\lambda\mu}\big\rangle = 
\bra{\Psi}  \hat{\Lambda}_l \; \hat{b}_{k\lambda\mu}\ket{\Psi} 
- \sum_{k'\lambda'\mu'}\frac{\mathcal{V}_{\lambda'}(k')}{\omega_B (k')}(\sigma^{\lambda'}_{\mu' 0})_l 
\bra{\Psi}\hat{b}^{\dagger}_{k'\lambda'\mu'} \hat{b}_{k\lambda\mu} \ket{\Psi}  
- \sum_{k'\lambda'\mu'}\frac{\mathcal{V}_{\lambda'}(k')}{\omega_B (k')}(\sigma^{\lambda'}_{0 \mu'})_l 
\bra{\Psi}\hat{b}_{k'\lambda'\mu'} \hat{b}_{k\lambda\mu} \ket{\Psi}\;.
\label{applying the shift to the hermitian conjugate}
\end{equation}
Again, $\bra{\Psi}\hat{b}^{\dagger}_{k'\lambda'\mu'} \hat{b}_{k\lambda\mu} \ket{\Psi}
\propto \delta_{\mu'0}\delta_{\mu 0}$, resulting in a vanishing contribution, similarly to
$\bra{\Psi}\hat{b}_{k'\lambda'\mu'} \hat{b}_{k\lambda\mu} \ket{\Psi}$. Therefore, we 
are left with 
\begin{equation}
\begin{aligned}
\big\langle \hat{\Lambda}_l\; \hat{b}_{k\lambda\mu}\big\rangle &= 
\bra{\Psi}  \hat{\Lambda}_l \; \hat{b}_{k\lambda\mu}\ket{\Psi} \\
& = |g_{LM}|^2 \bra{LM0}\bra{0}  \hat{\Lambda}_l\; \hat{b}_{k\lambda\mu} \ket{0}\ket{LM0} \\
& \qquad \qquad +g_{LM} \sum_{k'\lambda' n'} \alpha^{*}_{k'\lambda' n'} \bra{LMn'}\bra{0}
\hat{b}_{k'\lambda'n'}\;\hat{\Lambda}_l\;\hat{b}_{k\lambda\mu}\ket{0}\ket{LM0}\\
& \qquad \qquad \qquad+ g^*_{LM} \sum_{k''\lambda'' n''} \alpha_{k''\lambda'' n''} \bra{LM0}\bra{0}
\hat{\Lambda}_l\; \hat{b}_{k\lambda\mu}\;\;\hat{b}^{\dagger}_{k''\lambda''n''}
\ket{0}\ket{LMn''}\\
& \qquad \qquad \qquad\qquad + \sum_{k'\lambda' n'} \sum_{k''\lambda'' n''} \alpha^*_{k'\lambda' n'}
\alpha_{k''\lambda'' n''}\bra{LMn'}\bra{0} \hat{b}_{k'\lambda' n'}\;\hat{\Lambda}_l\;
\hat{b}_{k\lambda\mu}\;  \hat{b}^{\dagger}_{k''\lambda'' n''}
\ket{0}\ket{LMn''} \;.
\end{aligned}
\label{computing hermitian conjugate lambda b}
\end{equation}
The first two terms are evidently zero, since they involve an annihilation operator acting
on the phonon vacuum. The third term is trickier, but it ends up being proportional today
$\bra{0}\hat{\Lambda}_l \ket{0}$, reading a null contribution. Concerning the last term,
we have a different number of creation and annihilation operators, so it vanishes as well.

In the light of the calculations detailed above, we are left with 
\begin{equation}
\begin{aligned}
\bigg \langle \frac{d\hat{\pmb{\Lambda}}^2}{dt}\bigg\rangle &=
- i\sum_{l = \pm 1,0} \sum_{k\lambda\mu} (-1)^l \; \lambda(\lambda+1) \;
\mathcal{V}_{\lambda}(k)
\delta_{l,\pm 1} \delta_{\mu 0}\; \bigg(\braket{\hat{b}^{\dagger}_{k\lambda 0}} 
- \braket{\hat{b}_{k\lambda 0}} \bigg)\\
& = 2i \sum_{k\lambda} \lambda(\lambda+1) \; \mathcal{V}_{\lambda}(k)
\bigg(g_{LM}\alpha^*_{k\lambda 0} - g^*_{LM}\alpha_{k\lambda 0}\bigg) \\
& = -4 \sum_{k\lambda} \lambda(\lambda+1) \; \mathcal{V}_{\lambda}(k) \;\text{Im}\bigg(
g_{LM}\alpha^*_{k\lambda 0}
\bigg)\;,
\end{aligned}
\label{rearranging after expectation value}
\end{equation}
which is the final result reported in Eq. \eqref{final result after average}. Let us also remark
that, as a first consistency check, the above expression is real-valued, as expected from the 
expectation value of an observable.

%
%
\section{Time-dependent variational approach}
\label{app:3}

As mentioned in the main text, we consider here a time-dependent variational approach
based on the following Lagrangian functional \cite{kramer-book}, i.e.
\begin{equation}
\mathcal{L} = \bra{\Psi} \big(i\partial_t - \hat{\mathrm{H}} \big)\ket{\Psi}\;,
\label{lagrangian functional}
\end{equation}
where $\hat{\mathrm{H}} = \hat{U}^{-1}\hat{\mathcal{H}}_{\text{mol}}\hat{U}$, 
$\hat{\mathcal{H}}_{\text{mol}}$ being defined in eq. \eqref{hamiltonian molecular frame} and
$\hat{U}$ in eq. \eqref{unitary operator slow impurity}. In addition $\ket{\psi}$ is the 
one-phonon ansatz in Eq. \eqref{one-phonon ansatz}, encoding the time-dependent (complex) variational
parameters $g_{LM}$ and $\alpha_{k\lambda n}$. By symmetry \cite{lemeshko-2016},
we can also realize that the number of variational parameters we actually need is
reduced, since $\alpha_{k\lambda n} = \alpha_{k\lambda -n}$. For instance, in the
relevant cases where $\lambda = 0,\,1$ and $\lambda = 0,\, 2$, we need respectively 3 and 5 
equations.

The Euler-Lagrange equations are defined as 
\begin{equation}
\frac{d}{dt}\bigg(\frac{\partial \mathcal{L}}{\partial\dot{x}_i}\bigg) 
- \frac{\partial \mathcal{L}}{\partial x_i} = 0\;,
\label{euler-lagrange equation definition}
\end{equation}
with $\lbrace x_i \rbrace = g_{LM}, \alpha_{k\lambda}$. Now, deriving the actual Langrangian 
in terms of the variational parameters is a lengthy task, since we one needs to evaluate term by
term upon replacing $\ket{\psi}$ with Eq. \eqref{one-phonon ansatz}. Here, we refer to the
final result  
\begin{equation}
\begin{aligned}
\mathcal{L} & = i\bigg[g^*_{LM}(t)\dot{g}_{LM}(t) +\sum_{k\lambda n} \alpha^*_{k\lambda n}(t) 
\dot{\alpha}_{k\lambda\mu}(t)\bigg] + B\bigg[L(L+1) + \sum_{k\lambda}\lambda(\lambda+1) 
\frac{\mathcal{V}^2_{\lambda}(k)}{\omega_B^2(k)}\bigg]\; |g_{LM}(t)|^2  \\
& \qquad + \sum_{k\lambda\mu} \bigg[BL(L+1) +\omega_B(k)\bigg]\;|\alpha_{k\lambda n}(t)|^2
-2B\sum_{k\lambda\mu \nu} \pmb{\eta}^L_{n\nu} \pmb{\sigma}^{\lambda}_{\mu \nu}\; 
\alpha^*_{k\lambda n}(t)\alpha_{k\lambda\nu}(t)\\
& \qquad \qquad + B \sum_{kk'\lambda\lambda' n} 
\frac{\mathcal{V}_{\lambda}(k)}{\omega_B(k)}\; 
\frac{\mathcal{V}_{\lambda'}(k')}{\omega_B(k')} \; \delta_{n,\pm 1}\; 
\sqrt{\lambda(\lambda+1) \lambda'(\lambda'+1)}\;\alpha^*_{k\lambda n}(t)\alpha_{k'\lambda' n}(t)\\
& \qquad \qquad \qquad + B \sum_{k\lambda n} \frac{\mathcal{V}_{\lambda}(k)}{\omega_B(k)}\;
\delta_{n\pm 1}\;\sqrt{\lambda(\lambda+1)}\;g^*_{LM}(t)\alpha_{k\lambda n}(t)
+ B \sum_{k\lambda n} \frac{\mathcal{V}_{\lambda}(k)}{\omega_B(k)}\;
\delta_{n\pm 1}\;\sqrt{\lambda(\lambda+1)\;L(L+1)} \;g_{LM}(t)\alpha^*_{k\lambda n}(t)\;,
\end{aligned}
\label{lagrangian final result}
\end{equation}
where $\pmb{\eta}^L_{n\nu} = \bra{LM\mu}\hat{\mathbf{J}}'\ket{LM\nu}$, $\hat{\mathbf{J}}'$ being the
angular momentum operator acting in the molecular frame (cfr. Eq. \eqref{hamiltonian molecular frame}).
Consequently, the equations of motion read
\begin{equation}
- i \dot{g}_{LM} = B\bigg[L(L+1) + \sum_{k\lambda}\lambda(\lambda+1) 
\frac{\mathcal{V}^2_{\lambda}(k)}{\omega_B^2(k)}\bigg]\; g_{LM}(t)
+ B \sum_{k\lambda n} \frac{\mathcal{V}_{\lambda}(k)}{\omega_B(k)}\;
\delta_{n\pm 1}\;\sqrt{\lambda(\lambda+1)}\;\alpha_{k\lambda n}(t)
\label{equation of motion for gLM}
\end{equation}
and 
\begin{equation}
\begin{aligned}
-i \dot{\alpha}_{k\lambda n}(t) & =  \bigg[BL(L+1) +\omega_B(k)\bigg]\alpha_{K\lambda n}(t)
-2B\sum_{\nu} \pmb{\eta}^L_{n\nu} \pmb{\sigma}^{\lambda}_{\mu \nu}\;
\alpha_{k\lambda\nu}(t) \\
& \qquad +B \sqrt{\lambda(\lambda+1)} \;\frac{\mathcal{V}_{\lambda}(k)}{\omega_B(k)}
\sum_{k'\lambda'}\delta_{n\pm 1}
\sqrt{\lambda'(\lambda'+1)}\frac{\mathcal{V}_{\lambda'}(k')}{\omega_B(k')}\alpha_{k'\lambda' n}(t)
+B \sqrt{\lambda(\lambda+1)\; L(L+1)}\; \frac{\mathcal{V}_{\lambda}(k)}{\omega_B(k)}
\delta_{n\pm 1}g_{LM}(t)\;.\\
\end{aligned}
\label{equation of motion for alpha k lambda general}
\end{equation}
The simplest relevant cases are $\lambda = 0,\,1$ and $\lambda = 0,\,2$. In the first case, we
have 
\begin{equation}
\begin{cases}
-i \dot{\alpha}_{k00}(t) = \big[BL (L+1) +\omega_B(k)\big]\alpha_{k00}(t)\\
\\
-i \dot{g}_{LM} (t) = \bigg[BL(L+1) + 
2 \int_0^{\infty}dk\; \frac{\mathcal{V}^2_{1}(k)}{\omega_B^2(k)}\bigg]g_{LM}(t)
+2B\sqrt{2L(L+1)} \int_0^{\infty}dk\;\frac{\mathcal{V}^2_1(k)}{\omega_B^2(k)} \\
\\
-i \dot{\alpha}_{k10}(t) = \big[ BL(L+1) + \omega_B(k) \big]\alpha_{k10}(t)
-2B\sqrt{2L(L+1)}\alpha_{k11}(t)\\
\\
-i \alpha_{k11}(t) = \big[ BL(L+1) + \omega_B(k) \big]\alpha_{k11}(t)
-2B\sqrt{2L(L+1)}\;\alpha_{k10}(t)-2B\;\alpha_{k11}(t) \\
\\
\qquad \qquad \qquad \qquad \qquad+ 2B \frac{\mathcal{V}_1(k)}{\omega_B(k)} \int_0^{\infty} 
dk' \frac{\mathcal{V}_1(k')}{\omega_B(k')}\alpha_{k'11}(t)
+ B\; \sqrt{2L(L+1)}\frac{\mathcal{V}_1(k)}{\omega_B(k)} \; g_{LM}(t)\;.
\end{cases}
\label{equation of motion for lambda zero one}
\end{equation}
In the second case $\lambda = 0,\,2$, 
which holds, for instance, when a I$_2$ is immersed in $^4$He, one gets
\begin{equation}
\begin{cases}
-i \dot{\alpha}_{k00}(t) = \big[BL (L+1) +\omega_B(k)\big]\alpha_{k00}(t)\\
\\
-i \dot{g}_{LM} (t) = \bigg[BL(L+1) + 
6 \int_0^{\infty}dk\; \frac{\mathcal{V}^2_{2}(k)}{\omega_B^2(k)}\bigg]g_{LM}(t)
+2B\sqrt{6L(L+1)} \int_0^{\infty}dk\;\frac{\mathcal{V}^2_2(k)}{\omega_B^2(k)} \\
\\
-i \dot{\alpha}_{k20}(t) = \big[ BL(L+1) + \omega_B(k) \big]\alpha_{k20}(t)
-2B\sqrt{6L(L+1)}\;\alpha_{k21}(t)\\
\\
-i \dot{\alpha}_{k21}(t) = \big[ BL(L+1) + \omega_B(k) \big]\alpha_{k21}(t)
-B\sqrt{6L(L+1)}\;\alpha_{k20}(t)-2B\;\alpha_{k21}(t) \\
\\
\qquad \qquad \qquad -2B \sqrt{L(L+1)-2 }\; \alpha_{k22}(t) +
6\frac{\mathcal{V}_2(k)}{\omega_B(k)} \int_0^{\infty}dk' \frac{\mathcal{V}_2(k')}{\omega_B(k')}
\;\alpha_{k'21}(t) \\
\\
\qquad \qquad \qquad \qquad \qquad + B \frac{\mathcal{V}_{2}(k)}{\omega_B(k)}
\sqrt{6L(L+1)}\; g_{LM}(t) \\
\\
-i\dot{\alpha}_{k22}(t) = \big[ BL(L+1) + \omega_B(k) \big]\alpha_{k22}(t)
-2B\sqrt{L(L+1)-2}\;\alpha_{k21}(t)-8B\;\alpha_{k22}(t) \;.
\end{cases}
\label{equation of motion for lambda zero two}
\end{equation}
%
%
%
%
%
%

\twocolumngrid

\bibliography{molecular-torque}

\end{document}